\documentclass[twocolumn,amsmath,amssymb,pra,longbibliography,unsortedaddress]{revtex4}

\usepackage[pdftex]{graphicx} 

\usepackage{dcolumn}  

\usepackage{bm}       
 
\usepackage[usenames,dvipsnames]{color}

\definecolor{URLCOL}{rgb}{0,0.17,0.43} 

\definecolor{LINKCOL}{rgb}{0.05,0.4,0} 

\definecolor{CITECOL}{rgb}{0.35,0,0.48} 

\usepackage{epstopdf}

\usepackage[pdftex,bookmarks,breaklinks,bookmarksopen,bookmarksnumbered,colorlinks,
linkcolor=LINKCOL,linktocpage,citecolor=CITECOL,urlcolor=URLCOL,pdfpagemode=UseOutline,pdftex]{hyperref}

\usepackage{booktabs}

\usepackage{verbatim}

\usepackage{natbib}

\definecolor{TITLECOL}{rgb}{0.1,0.2,0.7} 

\definecolor{PCOL}{rgb}{0.5,0.06,0.01} 

\definecolor{CHAPCOL}{rgb}{0,0.48,0} 

\definecolor{SECOL}{rgb}{0.1,0.2,0.7} 

\definecolor{CONTENTSCOL}{rgb}{0.1,0.2,0.7} 

\definecolor{SSECOL}{rgb}{0.25,0,0.48} 

\definecolor{SSSECOL}{rgb}{0.2,0.08,0.53} 

\definecolor{SHDCOL}{rgb}{0.4,0,0} 

\definecolor{ITMCOL}{rgb}{0.4,0,0} 

\definecolor{EXCOL}{rgb}{0,0.47,0.01} 

\definecolor{DEFCOL}{rgb}{0,0.42,0.01} 

\def\coloredtitle#1{

 \newcommand*{\keeptitle}{#1}

\title{\textcolor{TITLECOL}{#1}}

} 

\def\coloredauthor#1{

\newcommand*{\keepauthor}{#1}

\author{\textcolor{CITECOL}{#1}}

} 

\def\sec#1{\section{\textcolor{SECOL}{#1}}}

\def\ssec#1{\subsection{\textcolor{SSECOL}{#1}}}

\def\sssec#1{\subsubsection{\textcolor{SSSECOL}{#1}}}

\def\bea{\begin{eqnarray}}

\def\eea{\end{eqnarray}}

\def\ben{\begin{equation}}

\def\een{\end{equation}}

\def\benu{\begin{enumerate}}

\def\enu{\end{enumerate}}

\def\bei{\begin{itemize}}

\def\eei{\end{itemize}}

\def\beit{\begin{itemize}}

\def\eit{\end{itemize}}

\def\benu{\begin{enumerate}}

\def\enu{\end{enumerate}}

\def\n{n}

\def\sss{\scriptscriptstyle\rm}

\def\hatH{{\hat H}}

\def\1var{(\bx_1...\bx\N)}

\def\half{\frac{1}{2}}

\def\br{{\bf r}}

\def\bx{{x}}

\def\s{_{\sss S}}

\def\xc{_{\sss XC}}

\def\Hxc{_{\sss HXC}}

\def\Hxcd{_{\sss HXC,dyn}}

\def\N{_{\sss N}}

\def\H{_{\sss H}}

\def\ext{_{\rm ext}}

\def\dn{_\downarrow}

\def\sph_int{ {\int d^3 r}}

\def\iota{a}
\def\kappa{c}

\begin{document}

\thispagestyle{empty}

\coloredtitle{Linear response time-dependent density functional theory
of the Hubbard dimer}

\coloredauthor{Diego J. Carrascal$^{1,2}$, Jaime Ferrer$^{1,2}$, Neepa Maitra$^3$ and Kieron Burke$^4$}
\affiliation{$^1$ Department of Physics, Universidad de Oviedo, 33007 Oviedo, Spain}
\affiliation{$^2$Nanomaterials and Nanotechnology Research Center, CSIC / Universidad de Oviedo, Oviedo,  Spain}
\affiliation{$^3$ Department of Physics, Hunter College, City University of New York, New York, NY 1006, USA}
\affiliation{$^4$ Department of Chemistry and of Physics, University of California, Irvine, CA 92697,  USA}
\email{dj.carrascal@gmail.com, ferrer@uniovi.es, nmaitra@hunter.cuny.edu, kieron@uci.edu}

\date{\today}

\begin{abstract}
The asymmetric Hubbard dimer is
used to study the density-dependence of the exact frequency-dependent
kernel of linear-response time-dependent density functional theory.  
The exact form of the kernel is given, and the limitations of the
adiabatic approximation utilizing the exact ground-state functional are shown.
The oscillator strength sum rule is proven for lattice Hamiltonians, and relative
oscillator strengths are defined appropriately.
The method of Casida for extracting oscillator strengths from a frequency-dependent
kernel is demonstrated to yield the exact result with this kernel. 
An unambiguous way of labelling the nature of excitations is given.
The fluctuation-dissipation theorem is proven for the ground-state exchange-correlation energy.
The distinction between weak and strong correlation is shown to depend on 
the ratio of interaction to asymmetry.
A simple interpolation between carefully defined
weak-correlation and
strong-correlation regimes yields a density-functional approximation for the kernel that gives accurate transition frequencies for
both the single and double excitations, including charge-transfer excitations.
Many exact results, limits, and expansions about those limits are given in the appendices.
\end{abstract}

\maketitle

\def\T{\hat{T}}

\def\V{\hat{V}}

\def\eps{\epsilon}

\def\deps{\Delta\eps}

\def\dn{\Delta\n}

\def\al{\alpha}

\def\gm{t}

\def\dv{\Delta v}

\def\MHtwo{^{\rm MH2}}

\def\MHtwor{^{\rm MH2,~Reg}}

\def\MHfour{^{\rm MH4}}

\def\MH6{^{\rm MH6}}

\def\WCtwo{^{\rm WC2}}

\def\WC4{^{\rm WC4}}


\sf

\sec{Introduction}

Time-dependent density functional theory (TDDFT) is a popular first-principles approach
to calculating low-lying optical excitations of molecules~\cite{RG84,Ullrichbook,TDDFTbook12}.  A typical
calculation first involves optimizing the structure within ground-state DFT
using some approximate exchange-correlation functional.  Then a linear-response
TDDFT calculation, usually solving RPA-type equations in frequency
space~\cite{C95a,C96,BA96,GPG00}, or via real-time propagation~\cite{YNIB06}, yields both transition frequencies and
oscillator strengths.  The TDDFT step almost always makes the adiabatic
approximation for the unknown and (generally) frequency-dependent
exchange-correlation (XC) kernel, in which its zero-frequency limit is used~\cite{M16}. This is simply the second functional-derivative of the 
exchange-correlation energy of ground-state DFT.
Usually, the same approximate XC functional is used for the first ground-state step and for the TDDFT step. 
Several thousand papers per year use this method to extract useful information
on electronic excitations, with typical transition frequency errors of order
0.25 to 0.5 eV~\cite{AJ13,JWPA09,EFB09,CH12,CH16}. 

However, in the three decades since the Runge-Gross theorem established the
formal exactitude of this approach\cite{RG84}, a variety of
situations have been identified where approximations fail, often qualitatively.
Among the most notorious are failures for charge-transfer excitations, whose
transition frequencies are typically grossly underestimated by the standard functionals~\cite{TAHR99,DWH03,T03,GB04c, M05c,M17}, 
but reasonable results can be obtained by using range-separated hybrids~\cite{TTYY04,SKB09,BLS10,KSRB12,KB14}.    Another one
is the complete absence of double-excitations from 
the spectrum within the adiabatic approximation~\cite{JCS96,TH00,MZCB04}.  Initial hopes of extracting
double-excitations from higher-order response theory were dashed by Ref. \cite{TC03,EGCM11}.  A simple
model of the frequency dependence for the specific case of a double excitation
close to one or a few single excitations  in a weakly correlated system~\cite{MZCB04,CZMB04}, is a useful tool
for a post-adiabatic TDDFT treatment called dressed TDDFT, and has been applied to a range of systems~\cite{MMWA11,MW09,HIRC11}
but has not been widely adopted.

While practical electronic structure calculations begin from the real-space
Hamiltonian, much useful insight and even semi-quantitative results can be
extracted from model Hamiltonians, especially when correlations are 
strong\cite{BLBS12,SK11}.
The paradigmatic  case in condensed matter is the (one-band) Hubbard model, which
is usually taken on an infinite lattice, and can be analyzed in 1-,2-, or 3 dimensions.
The model is characterized by only two parameters, a hopping energy between
nearest neighbours $t$ and an on-site Coulomb repulsion for doubly-occupied sites $U$, and site-occupation 
plays the role of the density.
Model Hamiltonians are not aimed at high levels of quantitative accuracy, but 
are designed to explore qualitative features of correlation physics.
For example, the 2D Hubbard model may
display the essential features of high-temperature superconductivity~\cite{Dc94,LNW06,A13}.

Thus, Hubbard (and more complex) chains have been used to study, e.g., correlation
effects in transport through single molecules and small quantum dots. 
They have also been used to explore full time propagation in TDDFT, going beyond the
linear response regime~\cite{B08,LU08,V08b,KSKVG10,T11,FFTAKR13,FT12,RP10,SDS13,FM14,FM14a,DSH18,KS18,KKPV13,KVOC11,MRHG14,TR14}.   It is usually 
relatively straightforward
to exactly solve the time-dependent Schr\"odinger equation in these cases.
It can also be easy 
to find the exact ground-state density functional ~\cite{RP08,RP10,CF12,CC13,FM14a}, and to propagate the fully
time-dependent Kohn-Sham equations within the adiabatically exact approximation,
in order to study its capabilities and limitations.

Interestingly, among all the papers using TDDFT in lattice models,
relatively few have studied frequency-domain linear-response
TDDFT (lrTDDFT) in interacting systems \cite{AG02,FM14,TR14}.  In the case of the two-site Hubbard dimer, 
Aryasetiawan and Gunnarson \cite{AG02}
did ground-breaking work in studying the performance of lrTDDFT for the symmetric dimer.
However, as emphasized in a recent review focussed solely on ground-state
DFT for the dimer~\cite{CFSB15}, many crucial DFT features can only be
seen when the dimer is made
asymmetric via a difference in the on-site potentials $\Delta v$ ~\cite{RP08,RP10}.
In fact one cannot really speak of density-functionals if restricting to symmetric cases, since there
is no dependence
on the ground-state density, as the site occupations
always remain identical.
In addition, the Kubo response of the asymmetric dimer shows two excitations, while only one
survives in the symmetric case.
Again, a few  recent works have noted this effect \cite{RP10}.

In the present article, we thoroughly explore the asymmetric
dimer within lrTDDFT, finding the exact non-adiabatic 
density-functional for the exchange-correlation kernel.  
In previous works~\cite{TK14,RNL13} the exact frequency-dependent kernel has been found for a given system: in Ref.~\cite{RNL13} an analytic expression is derived for a homogeneous two-electron density on a ring while in Ref.~\cite{TK14} a general numerical procedure is given for computing the kernel of a given system. This is, we believe, the first time that the exact frequency-dependent kernel as a functional of the ground-state density has been found for any model; the Hubbard model is simple enough to allow for  a complete analytic study. 
We
find that correlations are suppressed by asymmetry,
so that a weak correlation approximation remains accurate  even when
the ratio between the Coulomb repulsion  $U$ and the hopping integral $t$
is very large, as long as the asymmetry between the on-site energies $\Delta v$ is also large.  In fact, for sufficiently large
$\dv/U$, this weakly correlated kernel remains accurate, no matter how large $U/t$ is.
Only when $U$ is large relative to both $2t$ and $\dv$ does the weak-correlation kernel
fail.  Moreover, a simple expansion about the strongly-correlated limit, which we call the
Mott-Hubbard (MH) regime, suffices for all other cases, so that an appropriate interpolation
between the two yields accurate results for almost all parameter values.  Thus we
have found an accurate approximate kernel for both double and charge-transfer excitations,
that works in both weak and strong correlation regimes.
We note that this provides a useful explicit example of the frequency-dependence of the
kernel as a functional of the ground-state density for this model, but does not produce 
a general purpose density functional for this frequency dependence.

\begin{figure}[htb]
\includegraphics[width=\columnwidth]{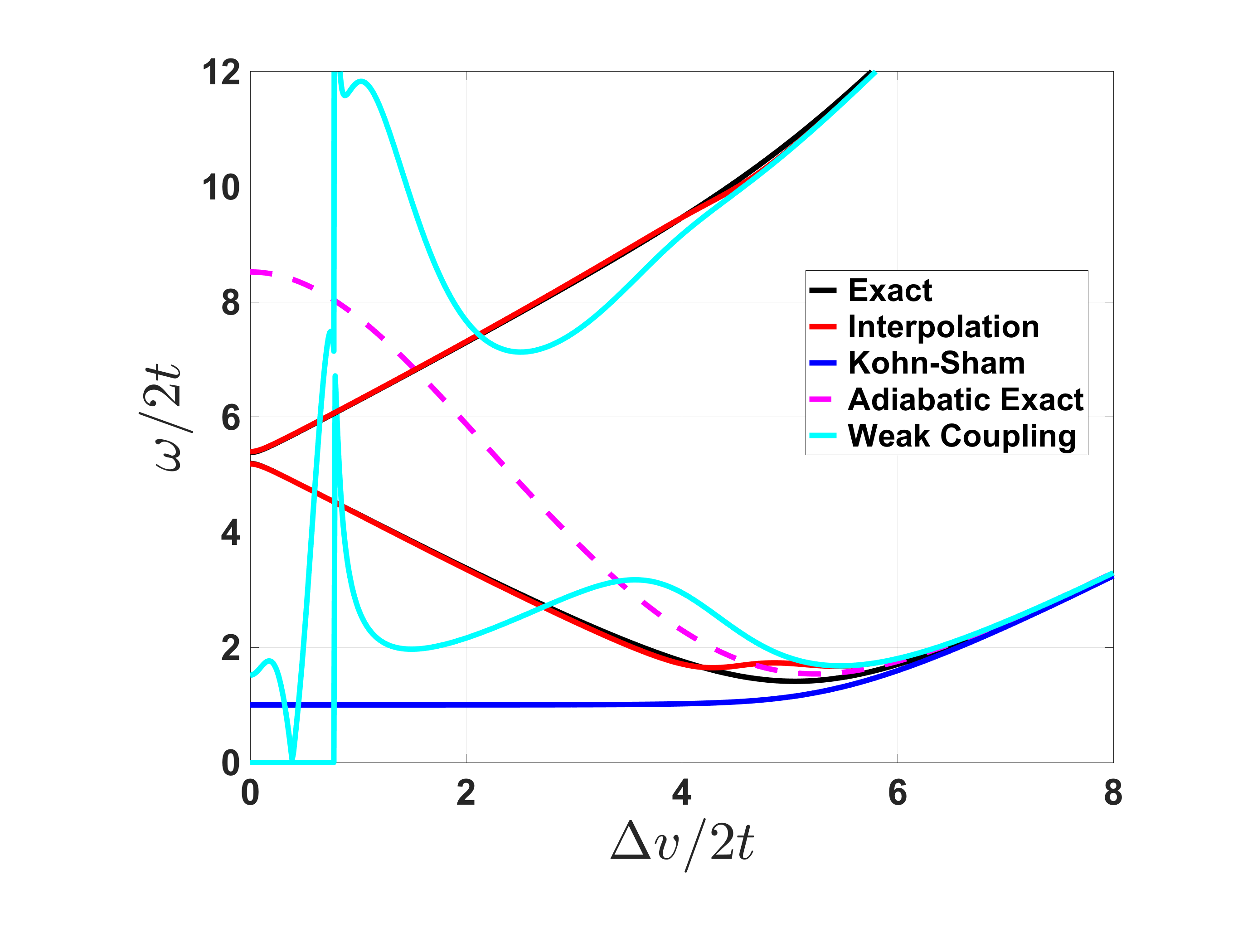}
\caption{Transition frequencies $\omega$ 
as a function of onsite potential asymmetry, $\dv$ for $U=10\,t$.
Black lines are exact, blue are the transitions of the
KS electrons with the exact ground-state functional, dashed magenta
includes TDDFT corrections with an adiabatically exact kernel, i.e.,
using the exact ground-state functional in TDDFT.
The cyan line shows TDDFT with a weak-correlation approximation
to the kernel, which diverges for sufficiently small asymmetry. 
The red line is the interpolation kernel developed in this work. 
Within this figure, 
the exact and interpolation lines can hardly be distinguished.}
\label{fig0}
\end{figure}

To illustrate these results we plot  in Fig. \ref{fig0} the transition frequencies
for both singlet excitations when the dimer is strongly-interacting ($U=10\,t$)
as a function of asymmetry, $\dv$.   In the symmetric limit ($\dv=0$),
the two excited states are barely separated.  Because correlation is strong,
the KS transitions are a poor approximation to the exact ones,
and even the adiabatically-exact
correction to TDDFT does not really improve matters.  It vastly
overestimates the correction to the single excitation and, being adiabatic,
yields no prediction for the double excitation at all.  The interpolation kernel developed
here, which interpolates between the weakly and strongly correlated limits, 
is almost perfect for these transition frequencies.
Note how, if the asymmetry is comparable to $U$ or larger, then the weak-coupling
approximation works well.  We explain this feature in this work.

While this article may appear long, its main results can be easily summarized.
In Section \ref{sec:tp4}, we give a very detailed account of how lrTDDFT behaves exactly
for the Hubbard dimer.  This is a beautifully simple case, with a very limited
Hilbert space, in which the (usually unknown) XC kernel of TDDFT can be
written exactly and explicitly (at least as function of the potential), including the 
frequency-dependence needed to
generate the double excitation.  This can be thought of as a many-body person's
guide to TDDFT.  On the other hand, in Section \ref{sec:sec4}, we explore meatier issues of
approximations.  We begin with weakly correlated systems (Sections \ref{sec:4b} and \ref{sec:tc11}) and
show how the usual approximations work in the usual way for such systems, drawing
the analogy with dressed TDDFT, which is a specific approximation to the frequency-dependent
kernel that captures double excitations in this regime.  But we also explore the
strongly-correlated (Mott-Hubbard) limit (Sections \ref{sec:4D} and \ref{sec:tc12}), and show how to distinguish weak
and strong correlation in this case.  We perform the necessary expansions in the
two limits (Appendix \ref{app:expansions}), and construct an interpolation scheme for the kernel that gives
highly accurate results in both regimes, and reasonably accurate results in
the interpolative regime (Section \ref{sec:tc13}).

For those with an interest and background in TDDFT, some key results to take away
include a general discussion of state labelling (how do you classify something
as a double excitation?: see Sections \ref{sec:mbt} and \ref{sec:ksr}), defining (relative) oscillator strengths 
in lattice models (see the same section and Section \ref{sec:ksr}), confirmation that the oscillator strength of
a double can be extracted from Casida's matrix formulation (Section \ref{sec:mat}), and illustration
that a pole in the kernel produces a double excitation, as in dressed TDDFT.
For those with a background in many-body theory, some other key results are
the separation of Mott-Hubbard and weakly correlated regimes (Section \ref{sec:4D} and Fig. \ref{fig:conu}),
and generalizations to site-dependent $U$ (Appendix \ref{app:U1U2}) and fractional particle numbers (Appendix \ref{app:fractional}).
The exact formulas (Appendix \ref{app:eigenvaules}) and expansions and limits (Appendix \ref{app:expansions}) should prove
very useful to anyone using Hubbard Hamiltonians with any background.

\begin{table}[htb]
\begin{tabular}{|>{$}l<{$}l|}
\toprule
\hline
\rm{Definition} & \rm{Description}\\ 
\hline
\midrule
\midrule
t & Hopping\\
U& Coulomb interaction, Hubbard U\\
v_1,\,v_2=-v_1            & On-site potentials \\
\Delta v = v_2 - v_1 & On-site potential difference\\ 
n_1,\,n_2 & Site occupations \\
N~{\rm or}~{\cal N} & Electron number \\
\Delta n=n_1 -n_2 & Occupation difference \\ 
\rho=|\Delta n|/2 & Useful alternative to $\Delta n$\\
|\Psi_i\rangle,\,E_i,\,\omega_i & Exact states, energies, and transitions\\
W_i & Exact excitation weights\\
f& relative oscillator strength of 2$^{nd}$ excitation\\
\mathrm{MHn}& Mott-Hubbard expansion to $n^{th}$ order\\
\mathrm{WCn}& Small-$U$ expansion to $n^{th}$ order\\
\midrule
\hline
\rm{Dimensionless}&\rm{variables}\\
\hline
\midrule
u=U/2t\,& Dimensionless Hubbard $U$\\
x=\Delta v/2t & Dimensionless potential difference\\
z=x/u & Potential diferrence in units of $U$ \\
\bar u,\bar x, \bar z & Reduced variables, ranging from 0 to 1\\
x_s=\Delta v_s/2t & Kohn-Sham potential difference\\
e_i=E_i/2t,\, \nu_i& Dimensionless energies and frequencies\\
\nu_3,\,\nu_4& Auxiliary frequencies\\
\chi& dimensionless response function\\
\chi_{AE} & Adiabatic approximation to $\chi$\\
a,\,b,\,c,\,\nu_f,& Exact response function parameters\\
f\Hxc & Hartree-exchange-correlation kernel\\
f_{st},\,f_{c,dyn}& Stationary and dynamic part of the kernel\\
\bottomrule
\hline
\end{tabular}
\label{notation}
\caption{Our Hubbard dimer notation. The dimensionless variables are set in units of twice the  hoppping 
unless otherwise stated. A subscript $s$ denotes a Kohn-Sham counterpart of any variable.}
\end{table}
Finally, we include here a table of notation that should help any reader dealing with the
many symbols used here.

\sec{Background}

\ssec{Time-dependent DFT}
\label{sec:tddft}
Time-dependent density functional theory (TDDFT) is based on the Runge-Gross theorem~\cite{RG84}, 
which is derived in a very different way from the Hohenberg-Kohn theorem of ground-state DFT\cite{HK64}.
The theorem proves a one-to-one correspondence between time-dependent densities and
one-body potentials, for a given initial-state, particle-particle interaction, and 
statistics.  Applied to electrons starting in a non-degenerate ground-state,
and using the Hohenberg-Kohn result that the ground-state wavefunction is a functional of the
ground-state density,  it implies that all properties of the many-body
system can be extracted from knowledge of its time-dependent density alone.

TDDFT can be and is applied to many-electron systems driven by arbitrarily strong
laser fields\cite{TDDFTbook12,M16}, but the vast majority of applications use results from linear-response.
Defining the density-density response function of a system as
\ben
\chi(\br,\br',t-t')=\left.\frac{\delta n(\br,t)}{\delta v\ext(\br',t')}\right\vert_{n=n_0(\br)},
\label{chidef}
\een
where $n_0(\br)$ is the ground-state density,
analysis leads to the famous Dyson-like equation \cite{GK85}: 
\ben
\chi(\omega)=\chi\s(\omega) + \chi\s(\omega)\star (f\H + f\xc(\omega))\star\chi(\omega),
\label{eq:chi-dyson}
\een
where $\star$
denotes matrix multiplication in $\br$-space (given two real-space functions, 
$f(\mathbf{r},\mathbf{r}')$ and $g(\mathbf{r},\mathbf{r}')$, matrix multiplication 
means $\int d \mathbf{r}''\,f(\mathbf{r},\mathbf{r}'')\,g(\mathbf{r}'',\mathbf{r}')$).
$\chi(\omega)=\chi(\br,\br',\omega)$ is the Fourier transform of
$\chi(\br,\br',t-t')$,
$\chi\s(\omega)$ is its non-interacting KS analog, 
while $f\H = 1/\vert\br - \br'\vert$ is the Hartree kernel 
and $f\xc(\omega) = f\xc[\n_0](\br,\br',\omega)$ is the
frequency-dependent XC kernel,  a functional of the ground-state
density. The latter is the  time-Fourier transform of $\delta v\xc(\br,t)/\delta n(\br',t')$. 
This Dyson-like RPA-like equation can in principle be solved for the exact 
$\chi$, which has poles at all optically-allowed excitations of the system.

For molecules, Eq.~(\ref{eq:chi-dyson}) is often re-cast in the form of
a matrix equation in the 
space of single KS excitations.  These can be derived and represented 
in several ways~\cite{C95a,GPG00,BA96,GM12}, but all are essentially equivalent
to finding eigenvalues and eigenvectors of the matrix
\ben
R_{qq'}(\omega) = \omega_q^2\, \delta_{qq'} + 4\sqrt{\omega_q\omega_{q'}}\, f\Hxc^{qq'}(\omega)\;,
\label{eq:casida}
\een
where 
\ben
f\Hxc^{qq'}(\omega)= [ \,q\vert f\H+f\xc(\omega))\vert q'\,],
\label{fHxc}
\een
and
$q=(i,a)$ represents a 
double-index, with $i$ labelling an occupied orbital and $a$ an unoccupied one, 
with

\ben
[\,q\vert f\xc(\omega)\vert q'\,]
=\int d^3r\, d^3r'\, \Phi_q(\br)\, f\xc(\br,\br',\omega)\, \Phi_{q'}(\br'),
\een
and $\Phi_q=\phi_i^*\phi_a$.
The eigenvalues of the matrix Eq.~(\ref{eq:casida})
are the squares of the transition frequencies $\omega_I$,
and oscillator strengths out of the ground-state, $f_I$, can be extracted from the eigenvectors.
The latter satisfy the Thomas-Reiche-Kuhn (TRK) sum rule \cite{T25,K25,RT25}
\ben
\sum_I f_I = N.
\een
In principle, both the transition frequencies and oscillator strengths are given
exactly when both exact ground-state and time-dependent DFT are used.
Even with the exact ground-state functional,
the KS response function has poles only at single excitations and, in the adiabatic
approximation, the excitations resulting from solving the matrix
equations yield only linear combinations of single excitations.
The frequency-dependence of $f\xc$ generates the states of multiple-excitation character. 

Practical DFT calculations require functional approximations.  In most applications
of TDDFT, the adiabatic approximation is made, allowing both the kernel and the
starting point to be approximated via ground-state functionals.  Such an approximation is
usefully accurate for many low-lying excitations of chemical interest~\cite{AJ13, JWPA09,EFB09, CH12, CH16}.
However, much experience has been gained on where standard semilocal approximations,
applied in this way, fail quantitatively or even qualitatively, including Rydberg
excitations, charge-transfer excitations, double excitations, conical intersections,
the thermodynamic limit, etc. \cite{M16}
More sophisticated functionals have been shown to offer a
good solution to several of these cases.  Many of these failures can be traced to errors made
in the ground-state part of the calculation;  these can be eliminated by using
the exact ground-state functional, when available, for simple model systems. 

Some years ago, a modest proposal was made for recovering double excitations
in lrTDDFT, at least in cases where the double was close to one or more single excitations,
and correlations were weak~\cite{MZCB04,CZMB04}.  By reverse engineering the exact wavefunctions in 
such a case, the frequency-dependent kernel of dressed TDDFT was proposed:
\ben
2[q\vert f\xc(\omega)\vert q ] =2[q\vert f\xc^A\vert q ] +\frac{\vert H_{qD}\vert^2}{\omega - (H_{DD} - H_{00})}
\label{fxcdoub}
\een
for the case of a KS single excitation $q = i \to a$ mixing with a KS double excitation $D$. 
Here, 
$f\xc^A$ is an adiabatic approximation to the kernel, and $H_{IJ}$ are matrix
elements of the full Hamiltonian between the KS states indicated.   
The additional pole in this kernel generates a double excitation
at approximately the correct transition frequency when the system is  weakly correlated.

\ssec{Asymmetric Hubbard dimer\label{sec:tp3}}
We analyse here the asymmetric Hubbard dimer model with two opposite-spin fermions:
\bea
\label{eq:tp2}
\hatH &=& -t\, \sum_{\sigma} \: (\hat{c}_{1\sigma}^{\dagger}\hat{c}_{2\sigma} + h.c) +
U \sum_{i} \hat{n}_{i\uparrow\,}\hat{n}_{i\downarrow} + \sum_i v_i \hat{n}_i\nonumber\\
&=&\hat{T}+\hat{V}_{ee}+\hat{V}_{ext}.
\eea
We set  $\bar{v}=(v_1+v_2)/2=0$ and rewrite the external potential term as 
$V_{ext}=-\Delta v\, \Delta n/2$,
where $\Delta{v}=v_2-v_1$ and $\Delta n=n_1-n_2$.
We use $2t$ to set the energy scale, and so define dimensionless measures of the interaction strength $u=U/2\,t$ and 
the asymmetry $x=\Delta v/2\,t$. The Hamiltonian has three basis states within the sub-space $N=2$, $S^2=0,\,S_z=0$,
so that it can be diagonalized analytically yielding a ground state and two excited states with energies $e_i$ and
wave functions $\Psi_i,\,i=0,\,1,\,2$. Explicit expressions are given 
in Appendix \ref{app:eigenvaules}.

The asymmetric dimer makes a beautiful illustration of all the principles of TDDFT, because
so many confusing features of TDDFT
have explicit formulas in this case due to the very small Hilbert space~\cite{CF12}.
A recent review of simply ground-state DFT using the asymmetric dimer references the
substantial literature on this~\cite{CFSB15}.   The density functional for fixed integer particle number $N$ is just a  function of the
site occupation difference $\Delta n$, and the KS system is just an asymmetric tight-binding problem. Explicit formulae for
fractional particle numbers ${\cal N}$ can also be drawn.
Many features, from the effect of strong correlation on the Green's function, to the
derivative discontinuity correction to the gap at integer $N$ , can be calculated exactly and often
explicitly.  While the XC energy functional cannot be written analytically, a parametrization
given in Ref. [\onlinecite{CFSB15}] is so accurate as to make no discernible error on the scale used here.
It can also be simply generalized to include distinct Coulomb energies on the two sites, and
so include the 2-site Anderson model as a special case (see Appendix \ref{app:U1U2}).

\sec{Linear response}
\label{sec:tp4}
For the present purpose, we must go beyond just ground-state properties, and
calculate the excited state energies and `optical' response.  We confine
ourselves to spin-conserving perturbations.  We emphasize that several 
results in this section already appear elsewhere, although not in
the forms presented here.

We will be interested in extracting information about excitations
in response to a weak perturbation.  
Define 
the dimensionless density-density linear response function, 
\bea
\tilde{\chi}(t,t')= \left(\begin{array}{cc}
\delta\Delta n(t)/\delta x(t')\vert_{\Delta n_0,N_0}&\delta N(t)/\delta x(t')\vert_{\Delta n_0,N_0}\\
\delta\Delta n(t)/\delta \bar{v}(t')\vert_{\Delta n_0,N_0}&\delta N(t)/\delta \bar{v}(t')\vert_{\Delta n_0,N_0}
\end{array}
\right)
\eea
However, $\hat{N}$ commutes with the Hamiltonian and we work in this article in a subspace with definite $N$ ($=N_0=2$). As a consequence,
only $\chi(t,t)=\delta\Delta n(t)/\delta x(t')\vert_{\Delta n_0,N_0}$ is different from zero and we drop henceforth the subindex $N_0$.
Nothing forbids choosing subspaces with non-definite value of $N$, it is just more complicated \cite{CF12}, and in this case the four matrix 
elements would be non-zero.
\ssec{Many-body theory}
\label{sec:mbt}
We work from now on with  $\chi(t,t)=\delta\Delta n(t)/\delta x(t')\vert_{\Delta n_0,N_0}$; whose Fourier transform with respect to $t - t'$ gives, in
the Lehmann representation \cite{mahanbook},
\ben
\chi(\nu)=\frac{2\,\nu_1\,W_1}{\nu_+^2-\nu_1^2} \,+\,\frac{2\,\nu_2\,W_2}{\nu_+^2-\nu_2^2},
\label{eq:chi_dimer}
\een
where $\nu_+=\nu+i\,\delta$,  $\nu=\omega/2t$ are dimensionless frequencies and 
the infinitesimal positive number $\delta$ enforces the causality of the response function and shifts 
the poles infinitesimally below the real axis. (Here $\chi$ is $2t$ times the dimensional
tight-binding version of Eq. (\ref{chidef})). The two excitations are
characterised by their frequencies 
and weights,
\ben
\nu_i=e_i-e_0,~~~
W_i = 
|\langle \Psi_{0}|\Delta\hat{n}|\Psi_i\rangle|^2.
\label{eq:S_def}
\een
whose explicit expressions are given in Appendix \ref{app:eigenvaules}.
We define as "first" and "second" excitations always  $\nu_1$ and $\nu_2$ respectively, e.g.: "first" is the  lower and 
"second" is the higher of the two excitations of the Hubbard dimer. 
The weight of the second excitation vanishes in the symmetric case: 
much of what can be learned about how TDDFT works for strongly
correlated cases requires asymmetry.
The frequency integral of the imaginary part of $\nu\chi(\nu)$ is (see Appendix \ref{app:sumrule}):
\ben
-\int_0^\infty\frac{d\,\nu}{\pi}\,Im\,\chi(\nu)\,\nu=
\nu_3=\nu_1\,W_1+\nu_2\,W_2.
\label{eq:sumrule}
\een
For a real-space Hamiltonian, this integral satisfies the TRK sum rule \cite{T25,RT25,K25}, where the
right hand-side is just $N$, and so can be used to define oscillator strengths.
Because of the lattice nature of the model, this rule is not true here \cite{M77,BGR87,K64}, and the
right-hand side is not a universal value, independent of the interaction or potential.
We define a relative oscillator strength for the second excitation as
\ben
f=\frac{\nu_2\,W_2}{\nu_1\,W_1+\nu_2\,W_2}=\frac{\nu_2\,W_2}{\nu_3},
\label{eq:f_def}
\een
so that Eq. (\ref{eq:chi_dimer}) can instead be written as
\ben
\chi(\nu)= 2{\nu_3}\left(\frac{1-f}{\nu_+^2-\nu_1^2} \,+\,\frac{f}{\nu_+^2-\nu_2^2}\right).
\label{chi}
\een
Throughout our analysis, we will also use an equivalent form, namely
\ben
\label{eq:chiinv}
\chi^{-1}(\nu)=\iota\,\nu_+^2-\kappa-\frac{b\,\nu_+^2}{\nu_+^2-\nu_f^2},
\een
where, defining
\ben
\nu_4=\nu_1\,W_2+\nu_2\,W_1,~~ 
\nu_f={\sqrt{\frac{\nu_1\nu_2\nu_4}{\nu_3}}},~~
\bar\nu_f=\nu_f \frac{\nu_3}{\nu_4},
\een
then, with $\Delta\nu=\nu_2-\nu_1$, $\Delta\nu_f=\bar{\nu}_f-\nu_f$:
\ben
a=\frac{1}{2\nu_3},~~
\frac{b}{a}=\Delta\nu^2-\Delta\nu_f^2,~~
\kappa=\frac{\nu_1\,\nu_2}{2\,\nu_4}.
\een
Thus the response can be characterized by four functions ($a,b,c$ and $\nu_f$)
of the basic reduced variables
$u$ and $x$, which can be deduced from Eqs. (\ref{eq:chi_dimer}), (\ref{eq:sumrule}), and (\ref{eq:chiinv}). 
We will consider many approximations to $\chi$, but all will have
the same form as the exact $\chi$ of Eq. (\ref{eq:chiinv}), and therefore can be defined in terms
of $a,b,c$, and $\nu_f$.

\begin{figure}[htb]
\includegraphics[width=0.9\columnwidth]{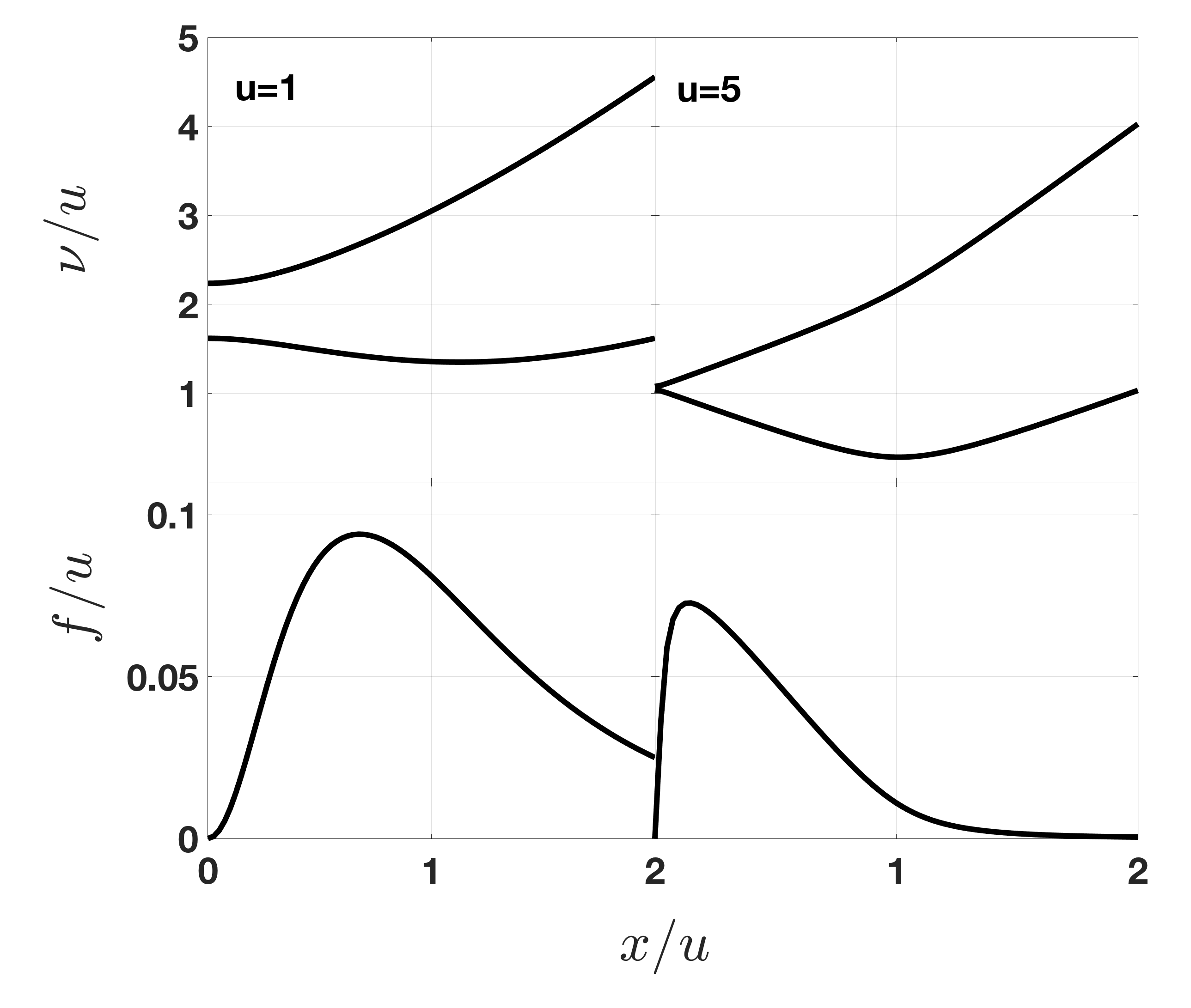}
\caption{Transition frequencies of the first and second excitations and oscillator strength of
the second excitation as a function of onsite potential asymmetry
$x=\Delta v$, for $u=1$ and $u=5$, where $2t=1$.}
\label{freq}
\end{figure}
Figure \ref{freq} shows the transition frequencies and relative oscillator strength $f$ (of
the second excitation) as a function of the dimensionless potential asymmetry $x=\Delta v/2t$
 for two different values of $u$.
On the left, $u=1$ and the system is weakly interacting. The first excitation
frequency initially drops with $x$, with the correction being $(1-3u)x^2/2$,
but eventually grows as $x$ when $x$ is larger than $u$.  The second
excitation has no linear
correction in $u$, and so behaves largely as its non-interacting value,
being $2$ in the symmetric case, and $2x$ for large $x$. 
The situation is very different when interaction is strong ($u=5$). Now, the frequency
of the two excitations equals about $u$ in the symmetric limit. These frequencies split
linearly however as $x$ grows as $u\pm x$ all along the Mott-Hubbard (MH) regime, that covers all
values of $x$ smaller than $u$ (see Sec. \ref{sec:4D}). This behaviour changes as soon as $x$ becomes larger than
$u$, where the system enters the Charge-Transfer (CT) regime. Subsequently, the 
frequency of the first excitation grows like $x-u$, while that of the second grows like
$2\,x$. The gap between the two hence grows linearly along the CT regime at a rate of $x+u$.
So we find that the excitations behave the same  for any value of $u$, for sufficiently large $x$. 
We will see later (Sec. \ref{sec:4D} )
that {\it sufficiently} means $x > u$, hence the CT regime. They however behave 
very differently for small and for large $u$ for small values of $x$, marking the MH regime $u > x$.   

It is useful to consider the nature of the ground and excited states in the extreme 
MH and CT limits to further understand these curves. Simplified expressions for the three states in these
limits can be found in Appendix \ref{app:wavefunctions}. In the MH limit of very large $u/x$ (i.e. towards 
the left of each plot in Fig. 2), the ground-state approaches one fermion on each site. This means the 
lowest excitation  transfers one fermion to the lower site, costing an energy of $u-x$, while the second 
excitation transfers one fermion to the upper site, costing an energy of $u+x$. On the other hand, in the CT 
limit of $u/x$ very small, the ground-state approaches the situation where both fermions sit on the lower site. The lowest 
excitation transfers one electron to the other site, costing an energy of $-u+x$, while the second 
excitation transfers both to the upper site, costing an energy of $2x$ relative to the ground-state. 
These limiting behaviors are evident in the plots above.

\ssec{KS response}
\label{sec:ksr}
In the previous section, we discussed our system within a traditional many-body
framework, with all parameters considered as functions of $u$ and $x=\Delta v/2t$, the 
interaction and one-body potential respectively.
This next section is devoted to showing how this system is treated exactly from
a TDDFT viewpoint, using the ground-state density in place of the one-body potential. 
Notice that we are working within the sub-space $N=2$. 
However, we write down analytical formulae for the KS response function for fractional occupation numbers ${\cal N}\in[0,4]$ in
Appendix \ref{app:fractional}. Knowledge of the dependence of the full response function on ${\cal N}$
relies on a complete analysis of the dependence of the XC kernel on ${\cal N}$, 
which is beyond the scope of
this article. The ground-state DFT analysis of the
Hubbard dimer for arbitrary integer or fractional ${\cal N}$ groundstate was discussed in detail in 
Ref.  [\onlinecite{CFSB15}].

The exact ground-state KS system is simply the asymmetric tight-binding model whose ground-state
site occupation difference matches that of the interacting system, i.e., $x\s(\rho) =\Delta v_s/2t = \rho/r$,
where $\rho=|\Delta n|/2$ is the {\em exact} interacting ground-state density and 
where $r={\sqrt{1-\rho^2}}$.
Thus it is trivial to construct the KS potential as a function of the ground-state
density.  The tight-binding model has two orbitals, the lower being doubly occupied
and the higher unoccupied in the ground state.  These fictitious KS
electrons have a response function
\ben
\label{eq:chi_s}
\chi_s(\nu)=\frac{2\,\nu_s\,W_s}{\nu_+^2-\nu_s^2},
\een
where 
\bea
\label{eq:tp1}
\nu_s=\sqrt{1+x_s^2}=\frac{1}{r},\,\,\,
 W_s=\frac{2}{1+x_s^2}=2\,r^2.
 \eea
Thus
\bea
\label{eq:chiinv_s}
\chi_s^{-1}&=&a_s\,\nu_+^2-c_s,
\eea
where $\iota\s=1/4r$ and $\kappa\s=1/4r^3$.
Notice that the KS pole corresponding to the second excitation has zero weight,
i.e., $f\s=0$, $b\s=0$. This expression for $\chi_s$ is generalized to fractional particle number in Appendix \ref{app:fractional}.

We end this section with a digression to give a general definition of the
nature of an excitation within TDDFT.  Our definition applies
whenever the exact KS ground-state wavefunction is a single Slater
determinant, but can easily be generalized beyond that.   
In such a case, the nature of an excitation of the KS system
is clear, e.g., a double excitation is a Slater determinant
with two electrons excited from their ground-state orbitals.
We note that 
the Hilbert space of states of the system is classified into subspaces
labelled unambiguously with  every set of quantum numbers available,
that includes $N$.
Then, the number of KS slater determinants and the number of exact 
many-body states in every subspace is the same.
As a consequence, each  many-body excitation can be continuously connected
to a well-labelled KS state via the adiabatic connection, i.e.,
by following its behavior as a function of $\lambda$, while
keeping the ground-state density fixed. 
This gives an unambiguous labelling to each level of the
many-body system.  This is the natural choice within KS DFT.
It differs from that of wavefunction theory, which usually
starts from the HF wavefunction.  The differences are small
for weakly correlated systems, but can be quite large when
correlation is strong.  In fact, when an unrestricted HF
calculation breaks symmetry, this creates difficulties in using
the HF wavefunction as a reference.  Here, the exact ground-state
KS wavefunction is always a doubly occupied singlet, and so 
does not suffer from this difficulty.

We follow this procedure here, and show, in strongly correlated
cases that, even when the interacting wavefunction is a mixture
of several determinants, its label remains unambiguous.  Of course,
when correlations are strong, the overlap between the many-body
and KS wavefunctions is often much less than 1, but this is also
true in the ground-state theory.
This definition must be applied carefully when curves
cross or in the thermodynamic limit, where there are infinitely
many states.
In Appendix \ref{app:wavefunctions}, we show how the many-body and KS 
states behave in the dissociation limit.  The adiabatic connection between the
many-body and KS wavefunctions can be traced down analytically in this limit, keeping the density fixed, and
so determine the nature of the wavefunctions, even though their overlap
at full-coupling is much less than 1.

\ssec{Exchange-correlation kernel}
From Eq. (\ref{eq:chi-dyson}), the Hartree-exchange-correlation kernel is defined by the difference of the true inverse response function 
from the KS inverse response function
\ben
\label{eq:chichis}
f\Hxc(\nu)=\chi\s^{-1}(\nu)-\chi^{-1}(\nu).
\een
This is in general a frequency-dependent quantity, but in almost all TDDFT calculations,
it is approximated by its static limit $f_{st}=f\Hxc(0)$.  For any finite system, this is exactly
given by ground-state DFT, and here
\ben
\label{eq:fst}
f_{st}=c-c\s.
\een
Moreover, with only two electrons, the exchange is precisely
minus half the Hartree, which has no frequency dependence.  Thus the interesting
dynamic contribution to the kernel is purely correlation,
\ben
\label{eq:fdyn}
f_{c,dyn}(\nu)
=(\iota\s-\iota)\,\nu_+^2+ \frac{b\,\nu^2}{\nu_+^2-\nu_f^2}.
\een 

\begin{figure}[htb]
\includegraphics[width=\columnwidth]{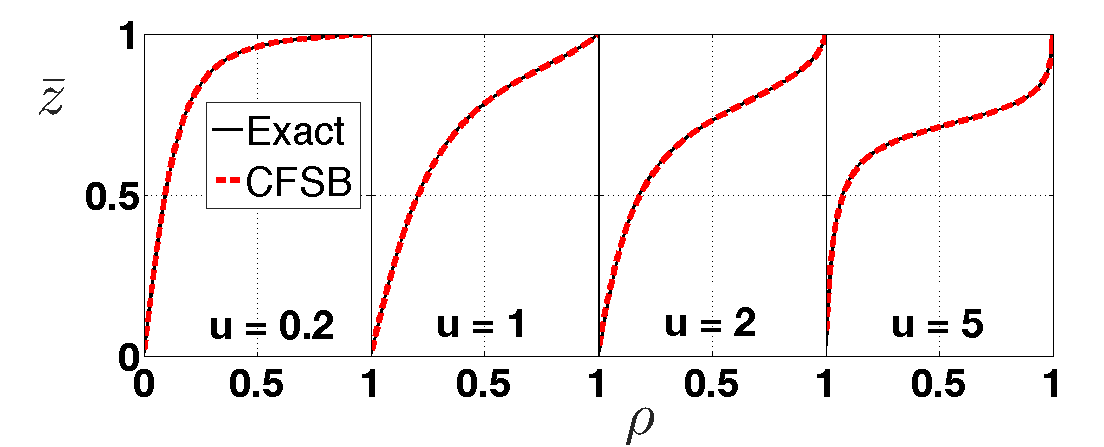}
\caption{Exact (black) and CFSB (dashed red) reduced external potential 
$\bar{z}=z/\sqrt{1+z^2}$, where $z=x/u$, as a function of 
$\rho=|\Delta n|/2$ for $u=$ 0.2, 1, 2 and 5 ($2t=1$).
}
\label{fig:zdn_cfsb}
\end{figure}

This dynamic contribution
depends on only three parameters, $\iota$, $b$ and $\nu_f$, which are in turn functions of
$u$ and $x=\Delta v/2t$. 
But, by virtue of ground-state DFT, the one-body potential is a unique function of the
density (difference), and so the three parameters in the kernel are functions of  $u$ and $\rho$,
which is how they appear in a TDDFT calculation.
This dependence can be found by using  the results of Ref. \cite{CFSB15} for the ground-state, 
that we  summarize in Appendix \ref{app:g0_review}.  In short, a very accurate
approximation for the universal contribution to the energy functional, ${\cal F}(\rho,u)$,
can be found. Since minimizing
the ground-state energy yields $x=-\partial {\cal F}/\partial\rho$, this is an explicit
expression for $x(\rho,u)$. This expression  can be inserted into the three parameters
to deliver the kernel functional. 
A comparison between the exact value of $z=x/u$ and the approximation is shown 
in Fig. \ref{fig:zdn_cfsb}, and any differences are invisible to the eye.

\begin{figure}[htb]
\includegraphics[width=\columnwidth]{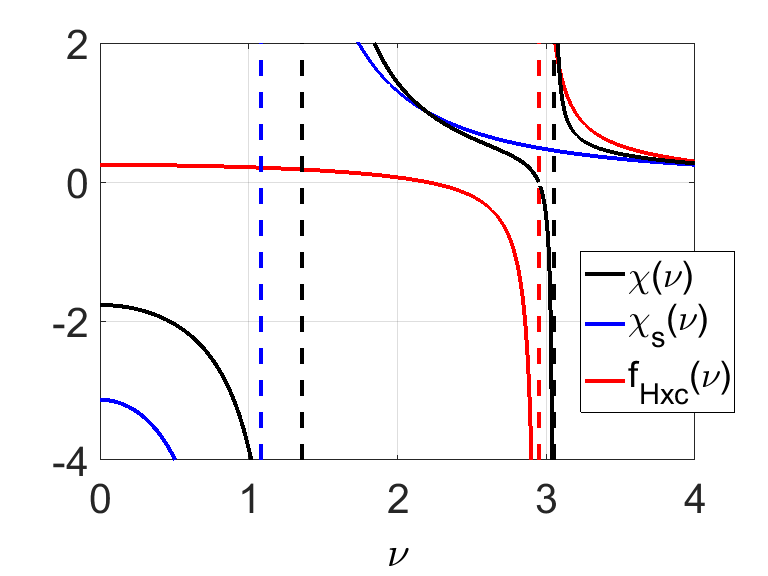}
\caption{Frequency dependence of exact (black) and Kohn-Sham susceptibilities (blue)
and exchange-correlation kernel (red line) for $u=x=1$, with
poles marked by dashed vertical lines, as a function of frequency $\nu$ ($2t=1$).
The red line shows the exchange-correlation kernel}
\label{chichi}
\end{figure}

In Fig. \ref{chichi}, we plot the response functions and kernel for $u=1$ and $x=1$, a relatively
weakly correlated and asymmetric system.  The exact response function (black) has poles at both the
first excitation ($\nu$ about 1.4) and the second (about 3.1).  The KS function (blue) has only
a single pole, corresponding to the KS first excitation, which is close to the exact first excitation
because this is a weakly-correlated case.  But there is no sign of the second excitation in 
the KS response.  The kernel has its own pole at about 2.95 which, when added to the KS
response function, produces the exact second excitation.  Note that this requires a pole
in the kernel at frequency $\nu_f$:  a smooth kernel would not produce the needed pole in $\chi$.  Note also that
expansions of the parameters in the kernel (in, for example, powers of $u$) do not yield
a well-defined expansion of the kernel itself, as they differ by arbitrarily large amounts for
frequencies near the poles.

In almost all applications of TDDFT, the adiabatic approximation
is used, i.e., $f\Hxc(\nu)$ is replaced by a constant.  We define
the adiabatically exact (AE) approximation
by replacing $f\Hxc(\nu)$
with the exact $f_{st}=f\Hxc(0)$ in Eq. (\ref{eq:chichis}). This yields
\def\AE{_{\rm AE}}
\ben
\chi\AE(\nu)=
\frac{1}{1/\chi\s-\,f_{st}}=\frac{2\,\nu_s\,W_s}{\nu_+^2-\nu\AE^2},
\een
where 
$\nu\AE=\sqrt{\nu_s\,\nu_1\,\nu_2\,W_s/\nu_4}$
is the (single) excitation frequency in the adiabatic approximation.
Since the AE approximation has no poles in the kernel, it fails to
generate any excitations in the response beyond those in the KS
response function, one of its principal
failings.  In fact, the weight and oscillator strength are identical to
the KS values.  It is simply that the position of the KS excitations
are shifted.  

\begin{figure}[htb]
\includegraphics[width=\columnwidth]{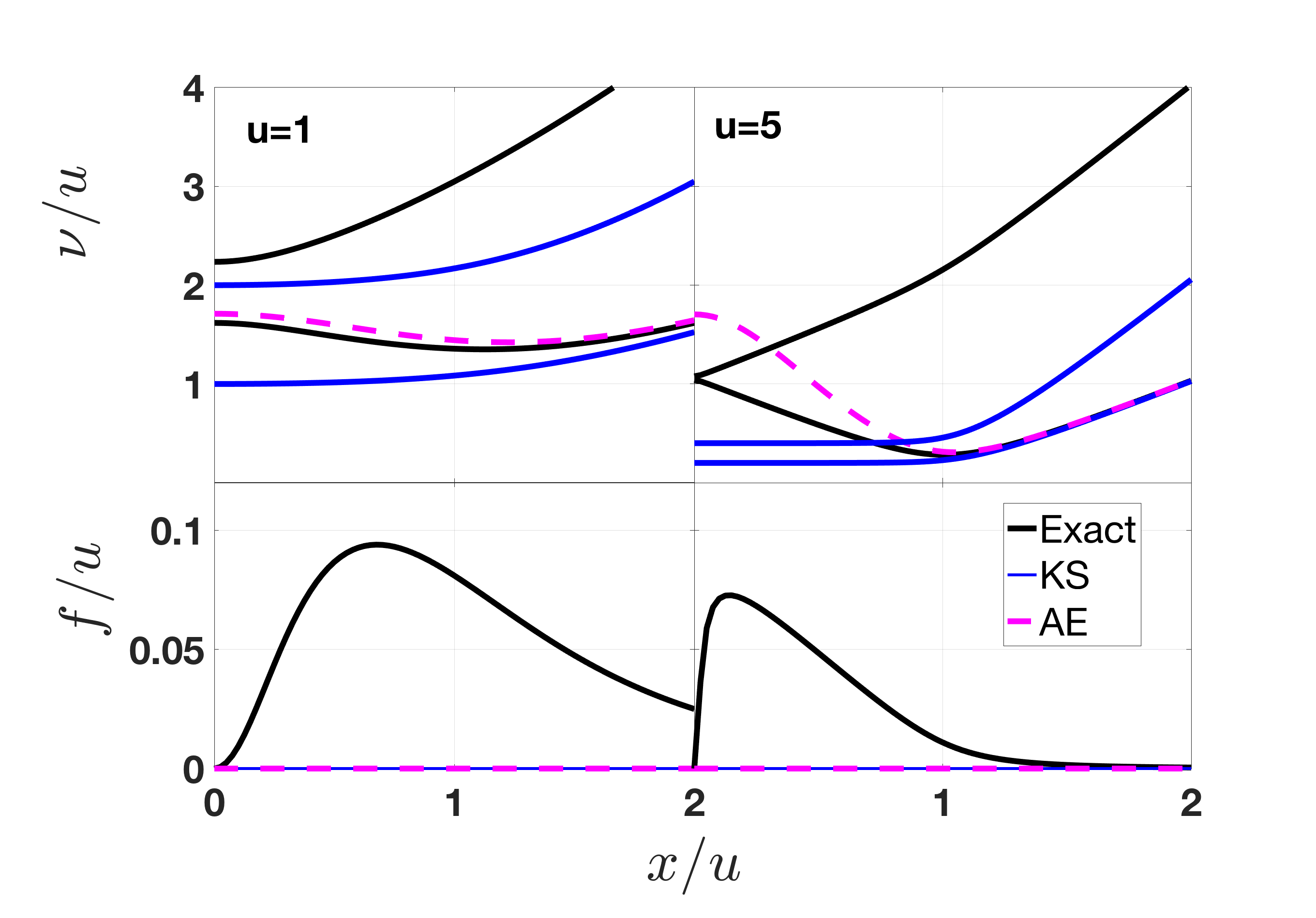}
\caption{Transition frequencies (top) and oscillator strength
for the many-body system (solid black), its Kohn-Sham counterpart (solid blue) and
the adiabatically exact approximation (magenta)
as a function of $x$, for $u=1$ (weakly correlated regime) and $u=5$ 
(strongly correlated regime) ($2t=1$). }
\label{freqs}
\end{figure}
In Fig. \ref{freqs}, we show the values of transition frequencies and
oscillator strength for both weak (left panel) and strong (right panel)
interaction.  For $u=1$ and in the $x/u>1$ domain for $u=5$, the KS values are a reasonable approximation
to the exact values, and the AE correction greatly improves the first
transition frequency.  In both cases (KS and AE), $f=0$, but the exact value of
$f$ is never greater than 0.1.  On the other hand, for $u=5$ and $x/u <1$, the KS single is a vast
underestimate relative to the exact single, the
AE is a serious overcorrection, the KS double (placed at double the KS single)
remains very far from its
physical value, and $f$ can be as large as 0.4, i.e., almost half
the oscillator strength can go into the second excitation. In the next subsection 
we will draw a close analogy between this behavior and that of a real stretched diatomic molecule. 
Thus a frequency-dependent kernel is vital to produce even qualitatively
correct excitations when correlation is strong.  Note that although
the first excitation improves when $x/u > 1$, and $f$ is also small,
the second transition
remains very badly described by its KS analog, even for high asymmetry.

When we come to discuss approximations to the dynamical
kernel, we will write these in terms of $a$, $b$, and
$\nu_f$.  The corresponding transition frequencies and oscillator strength
can be found directly from any such set.  Defining
$\gamma=(\nu_f^2+(\kappa+b)/a)/2$, and $\Delta=\sqrt{\gamma^2-\nu_f^2\,\kappa/\iota}$,
we find:
\ben
\label{eq:nu12W12a}
\nu_{1,2}^2=\gamma\mp\Delta,~~
f=\frac{1}{2}\,\left(1-\frac{\nu_f^2-\gamma}{\Delta}\right).
\een

We end this section with a well-known result.
In DFT, the fluctuation-dissipation theorem is often cited~\cite{LP75,GL76},
and can be the starting point of RPA-type approximations to the ground-state
XC energy.  In Appendix \ref{app:FD}, we show
\ben
\label{eq:tp11}
E\xc(\rho)=-\frac{U}{2}\,\int_0^1\,d\lambda\,
\int_0^\infty\,\frac{d\omega}{2\pi}\,Im\,\chi(\lambda U,\rho,\omega)-\frac{U\,N}{2}.
\een
This applies for either $N=1$ or $2$.
Here $\lambda$ multiplies $U$ everywhere, but $\rho$ is kept fixed.
This adiabatic connection is the DFT equivalent of the coupling
constant.  At $\lambda=1$, one gets the fully interacting system,
while at $\lambda=0$, the KS system is recovered, and 
$\chi^{\lambda=0}=\chi_s$.  Inserting our $\chi$ from Eq. (\ref{eq:chi_dimer}), 
we find a very simple form:
\ben
E\xc(\rho)=-\frac{U}{2}\,\int_0^1\,d\lambda\, \left(\sum_{i=1}^2 W_i(\lambda U,\rho)
-N \right),
\een
an elegant expression of the ground-state XC energy in terms of the weights of the
excitations.

\ssec{Matrix formulation}
\label{sec:mat}
The analog of the TDDFT matrix equation Eq.~(\ref{eq:casida}) for the Hubbard dimer is particularly simple due to the small Hilbert space. 
We can derive this from Eq.~(\ref{eq:chichis}), with the observation that $\chi(\nu)$ has a pole at the exact interacting frequencies $\nu_{1,2}$ (Eq.~(\ref{eq:chi_dimer})), and so $\chi^{-1}(\nu_{1,2})= 0$. Then, inverting Eq.~(\ref{eq:chi_s}) for $\chi\s^{-1}(\nu)$ on the right-hand-side of Eq.~(\ref{eq:chichis}), and rearranging  to solve for $\nu$, we obtain~\cite{FM14}
\ben
\nu^2 = \nu\s^2 + 2\nu\s W\s f\Hxc(\nu) \equiv R_{\rm H}(\nu)
\label{eq:casida_hub}
\een
whose solutions yield the exact frequencies of the interacting Hubbard dimer, $\nu_1, \nu_2$. This is the analog of what is known as the small matrix approximation for real molecules, when the matrix $R$ of Eq.~(\ref{eq:casida}) is truncated to just one single KS excitation. Since there is only one KS single excitation in the Hubbard dimer, Eq.~(\ref{eq:casida_hub}) is exact. 

As discussed in Sec.~\ref{sec:tddft}, oscillator strengths of real molecules are extracted from eigenvectors of the TDDFT linear response matrix Eq.~(\ref{eq:casida}). To obtain the oscillator strengths of the exact transitions in the Hubbard dimer from Eq.~(\ref{eq:casida_hub}), we retrieve 
a formula from Ref.~\cite{C95a}, which showed that the eigenvectors $G_I$ of the matrix Eq.~(\ref{eq:casida}) must be first normalized such that 
\ben
G_I^\dagger\left(1 - \left.\frac{\partial R}{\partial \omega^2}\right\vert_{\omega_I}\right) G_I = 1\;,
\een
before the oscillator strengths can be correctly extracted.
Since usually an adiabatic approximation is used, there is no frequency-dependence in the matrix $R$ and so this condition just reduces to requiring normalized eigenvectors. In fact, to our knowledge, there has not been any use of this result of Ref.~\cite{C95a} in the literature, likely because of the predominance of the adiabatic approximation. However, with a non-adiabatic kernel, such as we have in the Hubbard dimer, the frequency-dependence results in a rescaling of the eigenvectors, redistributing the oscillator strength in a way that depends on the excitation frequency. 
For our Hubbard dimer, this means the oscillator strength from the single KS excitation gets split into two, according to 
\ben
G_{1,2} = \frac{1}{\sqrt{1 - \left.\frac{\partial R_{\rm H}(\nu)}{\partial \nu^2}\right\vert_{\nu_{1,2}}}}
\een 
Taking the derivative of Eq.~(\ref{eq:casida_hub}), using Eq.~(\ref{eq:chi_s}) and Eq.~(\ref{eq:chi_dimer}) in Eq.~(\ref{eq:chichis}), then readily gives us 
\ben
G_i^2 = \frac{\nu_i W_i}{\nu\s W\s}\;.
\een
That is, the ratio of the transition strength of the second excitation to the total transition strength, is 
\ben
\frac{G_2^2}{G_1^2+ G_2^2} = \frac{\nu_2 W_2}{\nu_1 W_1 + \nu_2 W_2}\;,
\een
coinciding with our definition of $f$ in Eq.~(\ref{eq:f_def}). 

\sec{Weak and strong correlation}
\label{sec:sec4}
\ssec{Background}

Here we study the behavior of the system when interaction is weak, 
i.e., $u\lesssim 1$.  Of course, all quantities (excitation energies,
oscillator strengths, kernel parameters, etc.) can be expanded
as a power series in $u$, and the results are given in Appendix \ref{app:Useries}.
But we make a note of caution here:  There are many different expansions
in powers of $u$.  They differ in terms of which variable is held fixed.
From a many-body point of view, the natural expansion is holding the
external potential $x=\Delta v/2t$ fixed, and expanding in powers of $u$, which is the
meaning we have used so far.  However, even in ground-state DFT, the natural
expansion is the one used in the adiabatic connection formula, in which 
$\dn$ is held fixed.  This expansion differs from the many-body one.  As
we will show later, when dealing with strong correlation, even in many-body
theory, it will be more useful to hold the ratio $z=x/u$ fixed than keeping
$x$ fixed.

A second crucial point is that, in any of these expansions,
because of the frequency-dependence in the kernel and
the existence of a pole, there is no simple connection between
an expansion of the kernel parameters and the resulting behavior
of calculated transition frequencies.  Expansions in powers of $u$
do not commute with expansions in terms of the frequency, say.
It has long been known that, evaluating the kernel to leading
order in $\lambda$, i.e., at the exchange level, yields transitions
that contain {\em all orders} in $\lambda$, due to the non-linearity
of the RPA-type equation.  Thus, use of the exchange kernel leads to
approximate correlation corrections to the transitions.

\ssec{Relation to dressed TDDFT}
\label{sec:4b}
In the weak interaction limit the true excitations have a clear single 
and double excitation character respectively. Here we discuss some similarities and differences to dressed TDDFT.
First, dressed TDDFT isolates a single- and double-excitation
from among a spectrum of many excitations, assuming they are more strongly coupled to one
another than to any other.  Here, there are only these two excitations
in the entire spectrum.  This is why the
the exact kernel of Eq. (\ref{eq:fdyn}) has a simple pole of
the same type introduced in dressed TDDFT.  The only difference is 
that there are two poles here, $\pm\nu_f$,  which reflects the symmetric inclusion of
both forward and backward transitions.
However, the essential condition
of dressed TDDFT, namely that a specific single excitation is 
closest and most strongly coupled only to a specific double excitation, is not satisfied here.
For example, in the weak coupling limit, the double is at twice the 
frequency of the single, and no closer to it than the ground-state is.

\ssec{Weak-correlation kernel}
\label{sec:tc11}
To create an approximation that is appropriate for conditions
of weak correlation (corresponding to most current successful
applications of lrTDDFT), we expand in small $u$ for a fixed
value of $x$. We consider the
many-body expansion of Appendix \ref{app:Useries} in which we keep terms up to order $u^2$ in each of 
the parameters determining the kernel:
\bea
\label{eq:kernel2}
a^{WC2}&=&\frac{p_0}{4}\,\left(1-x^2\,\tilde{u}+\frac{(1+8\,x^2)\,\tilde{u}^2}{8}\right),\nonumber\\
b^{WC2}&=&\frac{9\,x^2\,p_0^3\,\tilde{u}^2}{16}
\,\left(1-\frac{2\,p_0^2\,\tilde{u}}{3}\right),\\
\nu_f^{WC2}&=&2\,p_0\,\left(1+\frac{p_0^2\,\tilde{u}^2}{8}\right).\nonumber
\eea
where $p_0={\sqrt{1+x^2}}$ and $\tilde u = u/p_0^3$, plus an extra term in the
expansion of $b$.   
With these expressions, we study the weakly correlated behaviour of 
the dimer.

\begin{figure}
\includegraphics[width=\columnwidth]{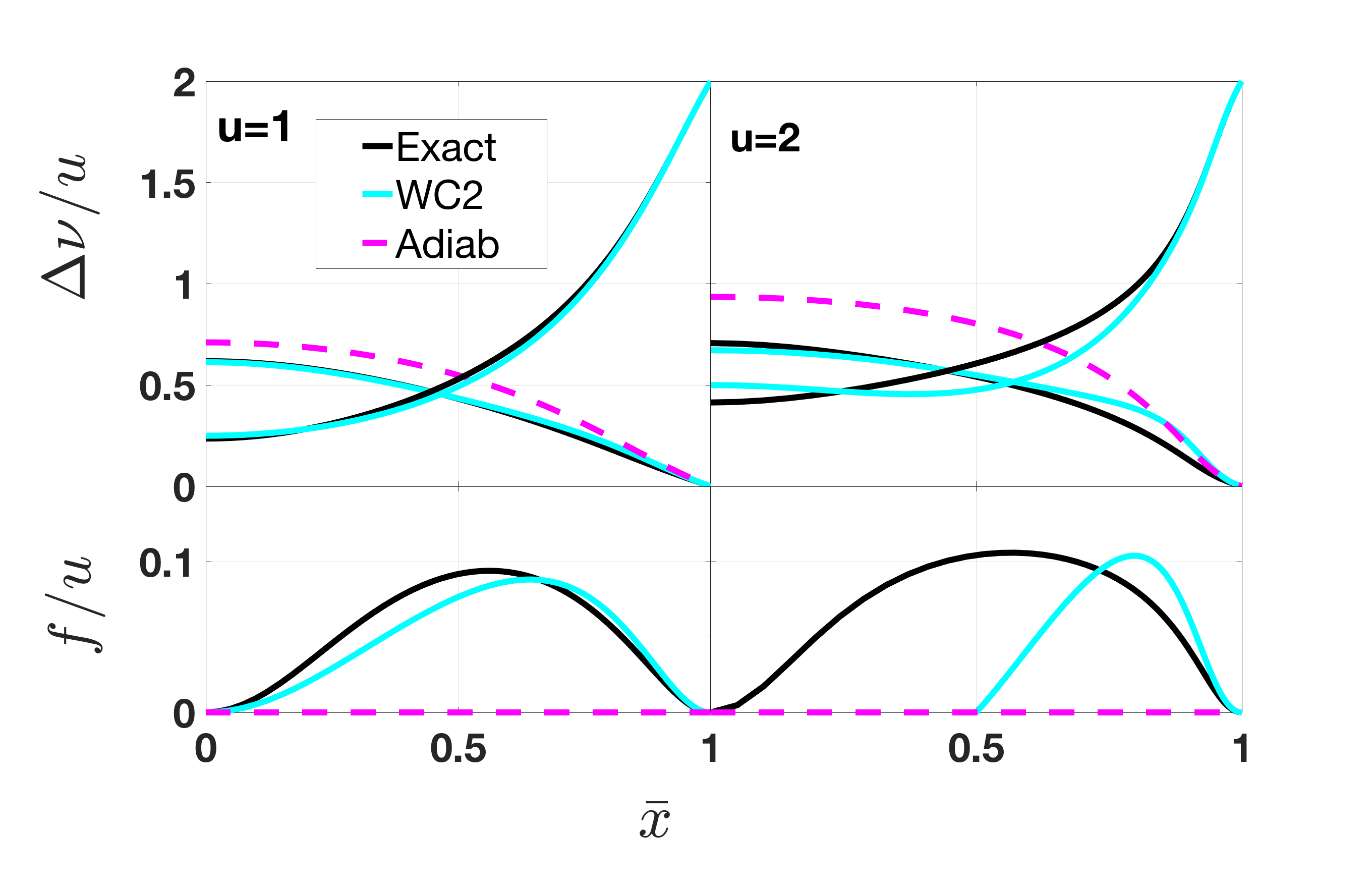}
\caption{Corrections to KS transition frequencies and oscillator strength as
a function of  $\bar{x}=x/\sqrt{1+x^2}$
for exact system (solid black), within AE approximation (dashed magenta) and
with the weak-correlation dynamical approximation
WC2 of Eq. (\ref{eq:kernel2}) (solid cyan). }
\label{WCres}
\end{figure}
In Fig. \ref{WCres} we plot the deviations of the transitions from their KS values $\Delta \nu_j=\nu_j-j*\nu_s,\,j=1,2$, both exactly
and for the AE and weakly-correlated approximations for $u=1$ and $u=2$.  We see that, in the weakly
correlated case ($u=1$ or less), the adiabatic
approximation for the transition frequencies is very close to the
exact quantity for both cases.  This is what is used (usually with
a ground-state approximation) in most applications of TDDFT.  However,
here we can also add the dynamical correction, expanded to leading
order in the strength of the correlation, and we find it improves
the results even further. This is especially apparent for the oscillator strength, where
the performance is very good, as $u=1$ is no longer very weak correlation.
However, once $u$ is large enough, this approximation must fail.
The weakly correlated approximation delivers poor results for the frequencies and
the oscillator strength for $u=5$, except for $\bar{x}=x/\sqrt{1+x^2}$ close enough to one.
We explore this point in the next section.

However, these are {\em not} explicit functionals
of the density, but rather they are post-calculation corrections to a standard
TDDFT calculation with an adiabatic kernel.
To convert them to density functionals, we express $x$ as a function of $\rho$ by using 
the relationship $x=-\partial f/\partial \rho$ and the ground-state density functional 
${\cal F}(\rho,u)$ described in Appendix \ref{app:g0_review}. We expand the functional  
in powers of $u$ as described in Appendix \ref{app:rhoWC} and find 
\bea
\label{eq:tp100}
x&\simeq&\frac{\rho}{r}+\rho\,u+
\frac{5}{8}\,\rho\,r^3\,u^2+
\frac{1}{4}\,\rho\,r^2\,\left(1-4\,\rho^2\right)\,u^3.
\eea
where $r={\sqrt{1-\rho^2}}$. This is then used to eliminate $x$ in Eqs. (\ref{eq:kernel2}) power by power, yielding:
\bea
\label{eq:kernelpu2}
a\WCtwo(\rho)&=&\frac{1}{4\,r}\,\left(1+\frac{1}{8}\,r^4\,u^2\right),\nonumber\\
b\WCtwo(\rho)&=&\left(\frac{3\,\rho\,r}{4}\right)^2 u^2\,\left(1+\frac{4}{3}\,r\,(1-3\,\rho^2)\,u\right),\\
\nu_f\WCtwo(\rho)&=&\frac{2}{r}+2\,\rho^2\,u+\frac{1}{4}\,r^3\,(1+9\,\rho^2)\,u^2\nonumber.
\eea

\ssec{When is a system strongly correlated?}
\label{sec:4D}
In this section, we discuss the concept of strong correlation in the
context of density functional theory, with special emphasis on the 
differences from many-body theory.  The key point is that, because the
exact KS system reproduces the exact density of the system, even when
correlations are strong, it can be a much closer mimic of the true
system than the traditional many-body starting point, namely a self-consistent
Hartree-Fock approximation, depending on what property is of interest.   
For example, when correlations are strong,
the lowest-energy self-consistent HF approximation breaks spin symmetry
(the unrestricted solution, UHF),
whereas the KS wavefunction always
remains a singlet, no matter how strong
correlation is (using the exact ground-state functional).
Thus the greatest differences occur just as correlations
become strong.

The first issue to address is how to decide when our dimer is strongly
correlated.  The most studied case is the symmetric case ($x=0$).  Here, 
it is clear that a Taylor expansion in small $u$ has a radius
of convergence of $u=2$ (branch cut at $u=2i$), while a similar expansion
in $1/u$ also converges up to $1/2$.  Thus $u=2$ is very definitively
the dividing point between weak and strong correlation.

But DFT is primarily concerned with inhomogeneous systems, which for our
dimer means asymmetry, so our definition must be generalized to all values
of $x$.  When the potential is highly asymmetric, does this categorization
change?  In fact, it does so, in an extremely important fashion.

\begin{figure}[htb]
\includegraphics[width=0.99\columnwidth]{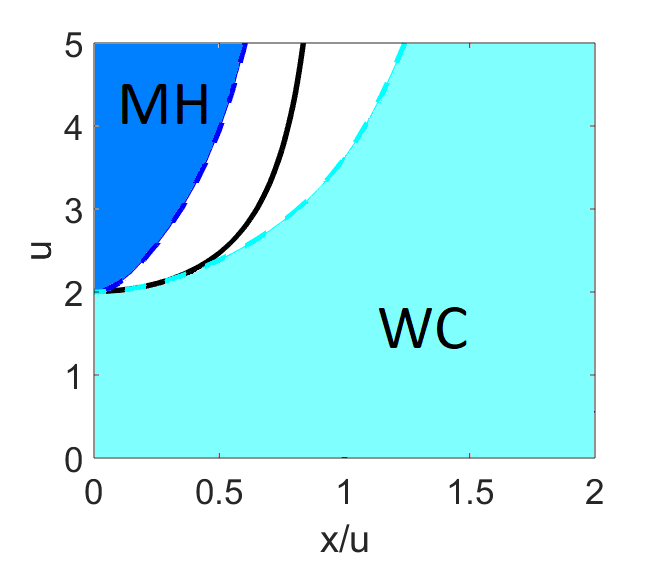}
\caption{Physical regimes in the Hubbard dimer:  Dark blue
is the pure Mott-Hubbard regime (limited error of approximations
around MH limit), while pale blue is the pure weakly correlated
regime (limited error of appoximation about the WC limit).  
The solid black line is the contour of 86\% overlap
between the many-body and Kohn-Sham wavefunctions.}
\label{fig:conu}
\end{figure}

In Fig. \ref{fig:conu}, we plot a contour of
the square overlap of the exact ground-state KS
wavefunction with the exact interacting wavefunction as a function of $\bar z$ and $\bar u$. 
We have chosen the value $\sqrt{3}/2 \approx 0.86$, as this yields precisely $u=2$ ($\bar{u}=1/\sqrt{2}$)
when $x=0$.  We have also colored in the region where Mott-Hubbard physics dominates (dark blue)
and the region where weak correlation approximations work (pale blue).   These will be quantified
below.  For now, the important lessons of Fig. \ref{fig:conu} are
first that {\em most} of the phase diagram is colored pale blue and
second that the variable on the x-axis is $x/u$, i.e., the asymmetry
divided by the interaction.  In fact, if this ratio is greater than 1, the
dimer is always weakly correlated, i.e., the black borderline
never crosses $x=u$, no matter how strong the interaction.
(The edge of the pale blue region simply delineates a contour of finite error for
the WC approximation, as described below).
This is because, in the ground state, both electrons sit on one site, despite
the strength of the interaction.

\ssec{Mott-Hubbard regime and expansions}
\label{sec:tc12}
\begin{figure}[htb]
\includegraphics[width=0.99\columnwidth]{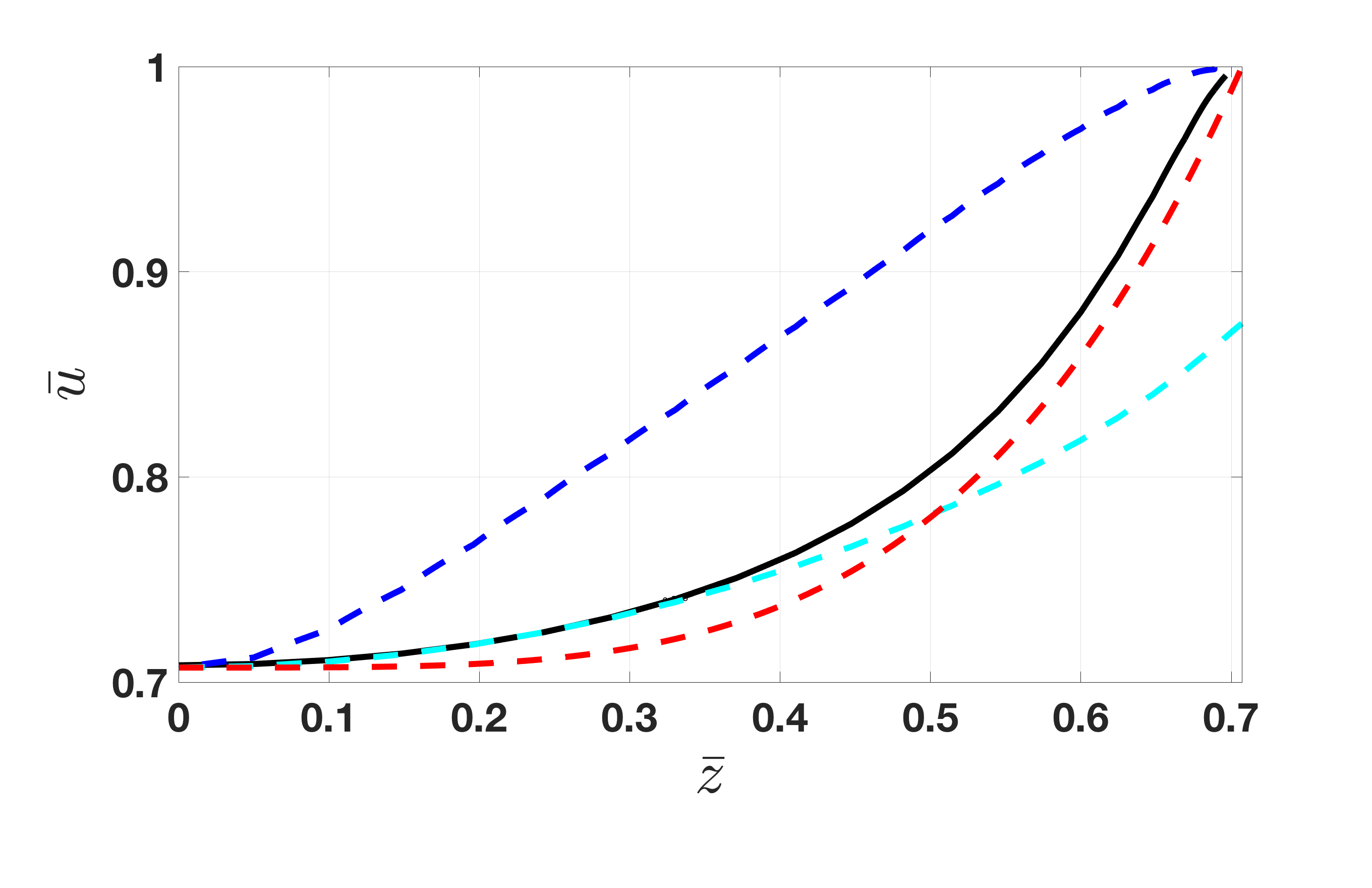}
\caption{Contour plot of the square of the overlap between the true and KS wavefunctions (black),
as well as contours
of error for WC2 (cyan) and MH approximations (blue), 
and our simple interpolation (red line).
}
\label{fig:contour}
\end{figure}

To 
capture the physics described above, we introduce a new variable
\ben
z = \frac{x}{u} = \frac{\dv}{U}.
\een
This is the onsite potential difference, but measured
on the scale of the interaction.  We show below that this is a more useful
variable than $x$ in considering strong correlation.
A similar variable was used in Ref.~\cite{LRG17} in their analysis of a Hubbard model of LiF. 
We also define the reduced
variables,
\ben
\bar u = \frac{u}{\sqrt{4+u^2}},~~~~\bar z = \frac{z}{\sqrt{1+z^2}}
\een
that run from zero to one as $u$ and $x$ span their whole range from zero to
infinity. Here, $u=2$ corresponds to $\bar u=1/\sqrt{2}$, while $x=u$ corresponds
to $\bar z=1/\sqrt{2}$. 

Figure \ref{fig:contour} replots Fig. \ref{fig:conu} in terms of the reduced
variables, and with more detail.  The solid black line is still the 86\%
overlap contour.
For $\bar{u}$ below this contour, the overlap  is at least this value,
and we consider the system weakly correlated.  
The first thing to notice is that the contour is confined to the
upper left corner of the $\bar u$-$\bar z$ plane.  In all the
remaining phase space, the overlap
is better than 0.93, including all $\bar z >1/\sqrt{2}$ (e.g.:$x>u$), 
{\em no matter how
large the value of $u$.}  It makes intuitive sense that for sufficiently
asymmetric systems, $u$ must be much larger to create strong correlation
effects.  What is notable is that the system is always 
weakly correlated when $x > u$.  This is the explanation for the success
of our weakly-correlated kernel to the right in the previous figures.

Now, the upper left corner (large $u$, small $x$) is the Mott-Hubbard regime, i.e., the
familiar physics of strong correlation in the symmetric limit.  In
this quadrant, the strong-correlation expansion described below is accurate.  
Above the blue contour, the strong correlation expression for the ground state energy has an error of  0.23 at most (in units of $2\,t$).
On the other hand, below the cyan contour, the WC2 approximation for the energy has
an error of only 0.086 at most (in units of $2\,t$ again).  The overlap contour runs neatly between these two.
Thus we need only the weakly-correlated and the MH regimes to cover all 
the physics in the dimer.  We can make a simple smooth interpolation to capture the contour, namely
\ben
\bar{u}_c(z) = a + b {\bar z}^p,
\label{eq:chempot}
\een
where $a$ and $b$ are positive real numbers, and $p$ a positive integer.
We find $p=4$ simulates the actual contour well.  Then $a=1/{\sqrt{2}}$
and $b=4(1-a)$ to achieve the correct limits.  This approximate contour
is also plotted in Fig. \ref{fig:contour}.

So, in order to capture the MH regime, we perform an expansion for large $u$, keeping
$z$ fixed and less than 1.  The results are
(Appendix \ref{app:tSeries}) 
\bea
\label{eq:MH}
a\MHtwo&=&\frac{u\,z_a}{8}\,\left(1+\frac{2\,(1+z^2)}{z_a^2 u^2}\right),\nonumber\\
b\MHtwo&=&\frac{u^3\,z^2\,z_a^2}{2\,z_b}\,\left(1+\frac{7\,z^4+18\,z^2-1}{z_a^2\,z_b u^2}\right),\\
\nu_f\MHtwo&=&u\,z_b^{1/2}\,\left(1-\frac{z^2-2}{z_a\,z_b u^2}\right).\nonumber
\eea
where $z_a=1-z^2$ and $z_b=1+3\,z^2$.   Clearly, these expressions fail for $z=1$ or larger, with
higher-order terms diverging.  The complementary expressions for $z > 1$ are the CT approximation,
and are given in the same appendix.

\begin{figure}
\includegraphics[width=\columnwidth]{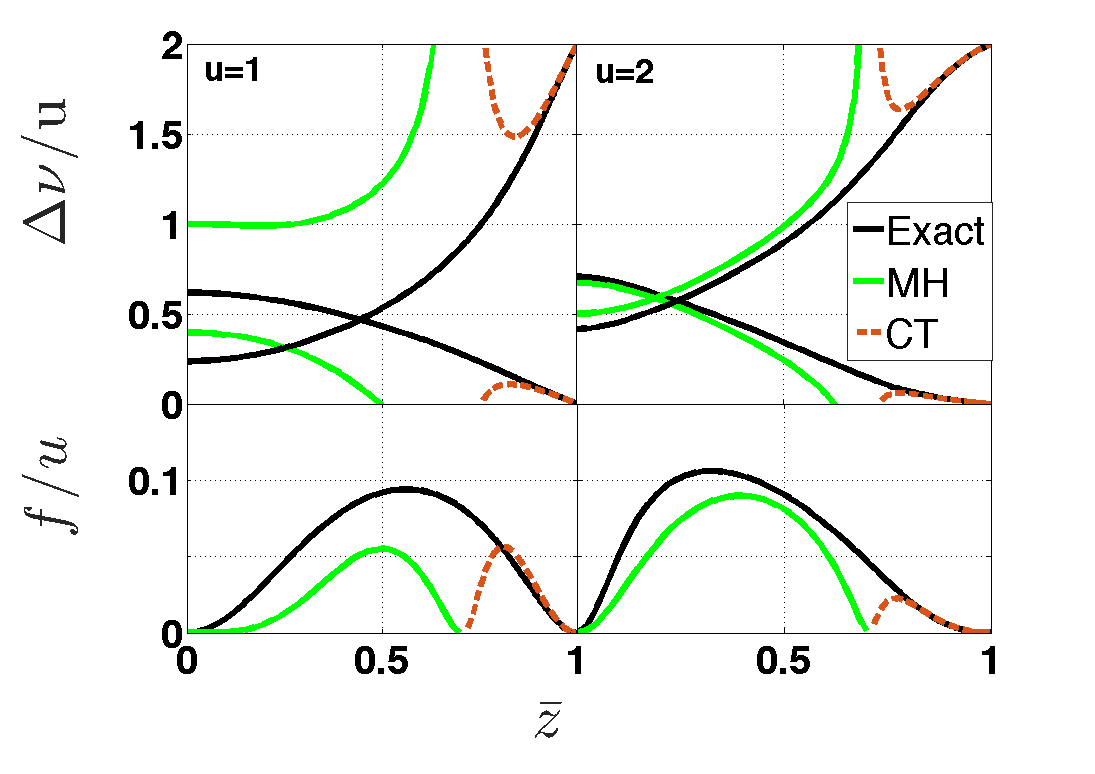}
\caption{Corrections to KS transition frequencies and oscillator strength as a function
of $\bar{z}=z/\sqrt{1+z^2}$ for exact
system (solid black) and with
the  MH2 (solid green) and CT (dashed brown) expansions.}
\label{fig:comparison_largeu_z}
\end{figure}
Figure \ref{fig:comparison_largeu_z} shows the exact deviations from the KS frequencies and 
oscillator strength alongside the MH and CT approximations.  
For larger $u$, MH works well until close to $x=u$, and CT works beyond that.
But clearly, near $z=1$, neither work well, and in fact diverge.  The region
in which this failure occurs shrinks with increasing $u$, but always exists.
For smaller $u$, such as $u=1$, this region is so large that the MH approximation
essentially never works, and CT only works for very large $z$.

\begin{figure}
\includegraphics[width=\columnwidth]{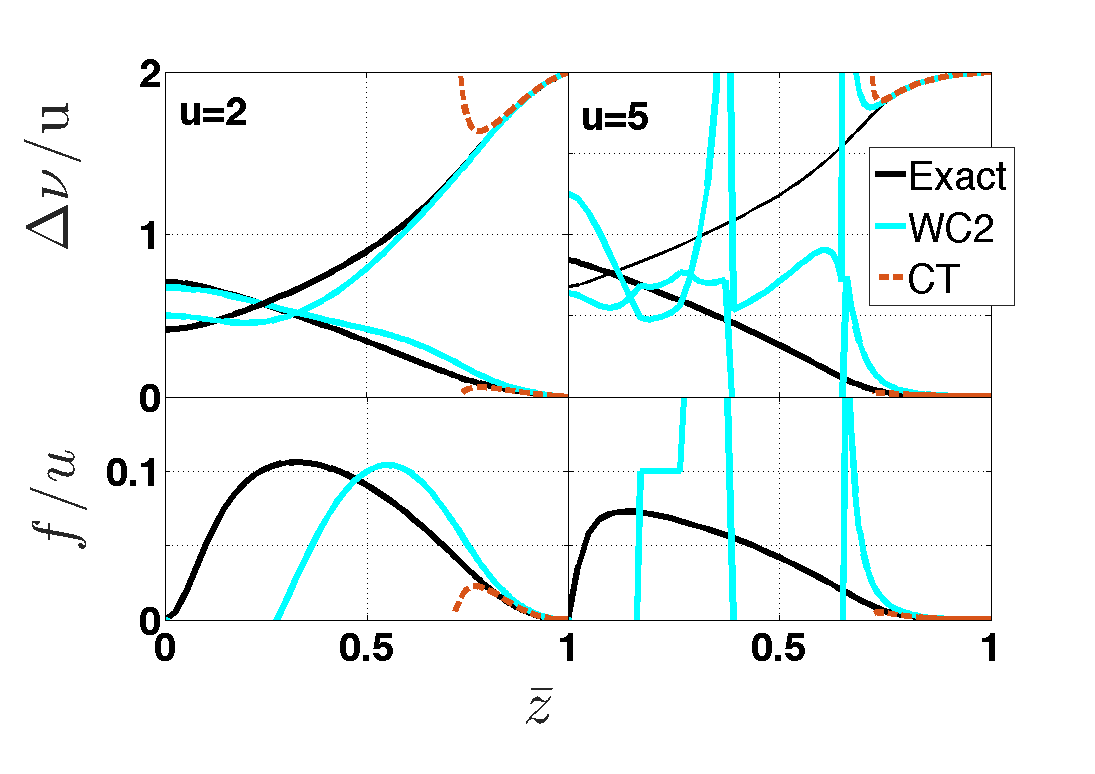}
\caption{Corrections to KS transition frequencies and oscillator strength as a function
of $\bar{z}$ exactly (black), WC2 (cyan) and 
CT expansion (dashed brown).}
\label{fig:comparison_weak_CT}
\end{figure}
In Fig. \ref{fig:comparison_weak_CT}, we compare the performance of the WC2 and CT expansions.  
For $u=2$ (left panel) and smaller, it is clear that WC2 is about the same
as CT for large $z$, but works much better for smaller $z$.  Even for $u=5$,
where WC2 fails badly for $z < 1$, it still works better than CT for $z > 1$.
In fact, we found no region in parameter space where CT outperformed WC2.
This is consistent with the contours of Fig. ~\ref{fig:contour}.

\ssec{Interpolation kernel\label{sec:tc13}}

\begin{figure}
\includegraphics[width=\columnwidth]{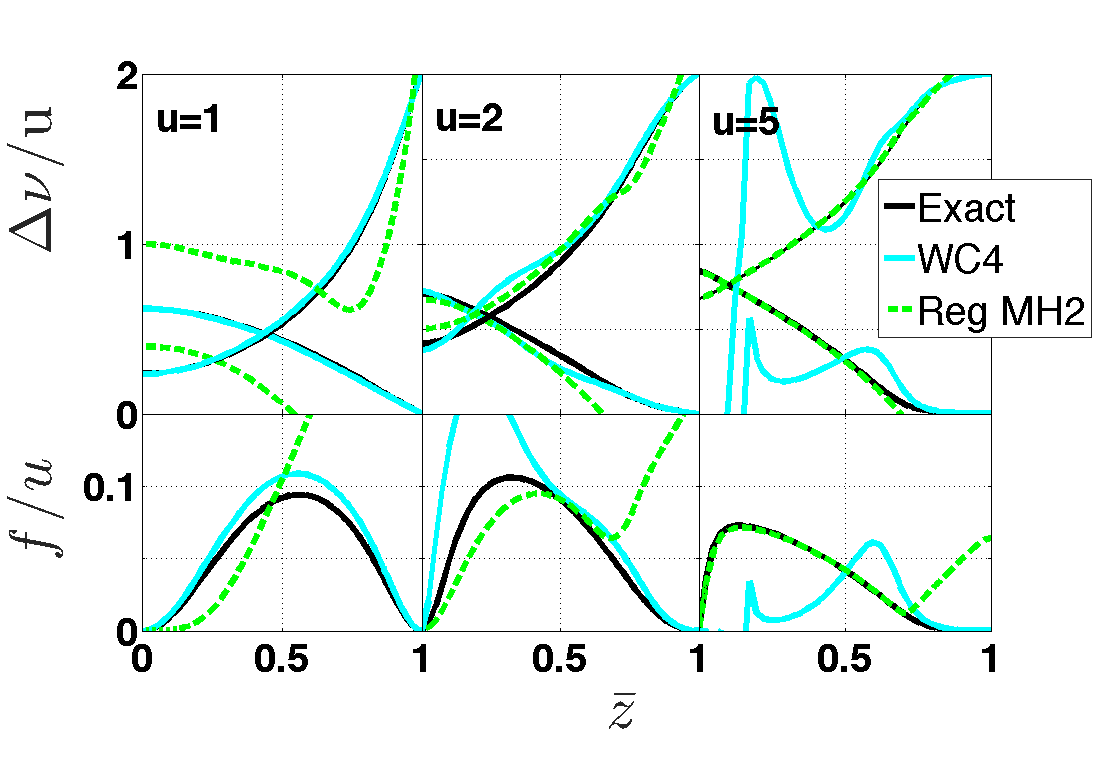}
\caption{Corrections to KS transition frequencies and oscillator strength as a function
of $\bar{z}$ exactly (black),
weakly correlated expansion WC4 (cyan) and regularized MH2 expansion (green).}
\label{fig:regularized}
\end{figure}

In this section, we construct an interpolation kernel between the MH and WC regimes.
We first improve the weakly correlated and MH approximations so that they 
match as smoothly as possible in the crossover region. 
We define WC4 as the expansion of the dynamic kernel parameters ($a$, $b$, and $\nu_f$)
to 4-th order in $u$, for fixed $x$.  The corrections to WC2 (Eq. )\ref{eq:kernel2})) are:
\bea
\label{eq:kernel4}
\Delta a\WC4&=&\frac{p_0\,\tilde{u}^3}{16}\,\left(x^2\,(4\,x^2-1)+\frac{16\,x^4\,(8\,x^2-9)-1}{32}\,\tilde{u}\right),\nonumber\\
\Delta b\WC4&=&\frac{p_0^3\,x^2\,(8\,x^4+58\,x^2+23)}{128}\,\tilde{u}^4,\\
\Delta\nu_f\WC4&=&\frac{p_0^3\,\tilde{u}^3}{4}\,\left(x^2+\frac{16\,x^4-9\,x^2-1}{16}\,\tilde{u}\right).\nonumber
\eea
We see in Fig. \ref{fig:regularized} that these clearly
improve the frequencies and
oscillator strength over WC2.
On the other hand,  while
adding one or two further terms in the MH expansion 
does not seem to improve matters much, 
removing divergences at $x=u$ does improve things.
We can regularize the MH2 expressions by replacing $u\,z_a$ with $\sqrt{u^2\,z_a^2+z^2}$.  This
provides a significantly smoother matching with the WC4 approximation at the crossover region when the 
interpolation scheme explained below is deployed. 
Figure \ref{fig:regularized} shows the impact of these
two schemes on the frequencies and oscillator strength, where we use
Eq. (\ref{eq:kernel4}) for the weak-coupling expansion
and (a regularized) Eq. (\ref{eq:MH}) for the MH expansion.
For $u=5$, we clearly see (regularized) MH working well up to $x=u$,
and WC4 working well beyond that (and each one failing outside its domain).
As $u$ is reduced, the regime where WC4 fails shrinks ($u=2$), until
for $u=1$, WC4 is almost perfect everywhere.

\begin{figure}
\includegraphics[width=1.0\columnwidth]{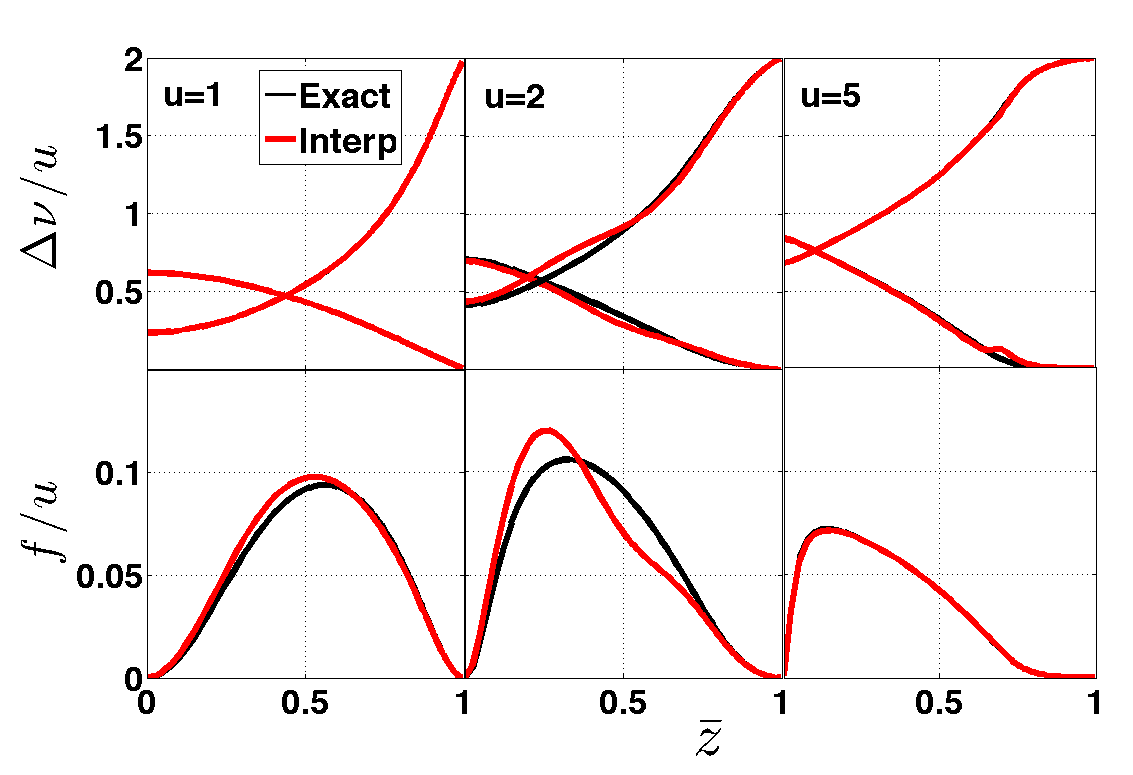}
\caption{Exact (solid black) and interpolated (red) deviations from KS frequencies and oscillator strength as a 
function of $\bar{z}$.}
\label{fig:interp_frequencies}
\end{figure}

We suggest the following interpolative scheme for each of the
kernel parameters:
\ben
\label{eq:interpolation}
a_{int}(u,z)=n_F\,a\WC4(u,z)+(1-n_F)\,\tilde a\MHtwo(u,z)
\een
where the tilde indicates that MH2 has been regularized, and
$n_F(u,z)$ varies smoothly from 1 to 0 as the contour $u_c(z)$ given by 
Eq. (\ref{eq:chempot}) (and shown in Fig. \ref{fig:contour})
is crossed.  We choose a Fermi function:
\ben
n_F(u,z)=\frac{1}{e^{\beta(u-u_c(z))}+1}.
\een 
We find $\beta=20$ yields a reasonably accurate transition.

We plot the results of the interpolation kernel for several values of $u$
as a function of $z$ in Fig. \ref{fig:interp_frequencies}.  
We see that it works reasonably well for $u < 1$ for
all $x$, and for $z > 1$ ($\bar{z}>1/\sqrt{2}$) for any $u$, and gives an imperfect but reasonable
interpolation in between.  This approximate kernel is not designed to yield the
extreme accuracy of the ground-state approximations of Ref. \cite{CFSB15}, but
just to show that once the limiting physics is included, an approximation can
be generated that works reasonably in all regimes.
Its limitations are most easily understood by starting with $u=5$,
where the error in the stitching is visible at $\bar{z}=1/\sqrt{2}$ (i.e.: $z=1$, or $x=u$), but it is small
and spans a relatively small region of $z$.  As $u$ is reduced, this region
grows, and is largest for $u=2$.  By the time $u=1$, this region has vanished
entirely, and the WC4 formula dominates and works well everywhere.

\begin{figure}
\includegraphics[width=1.0\columnwidth]{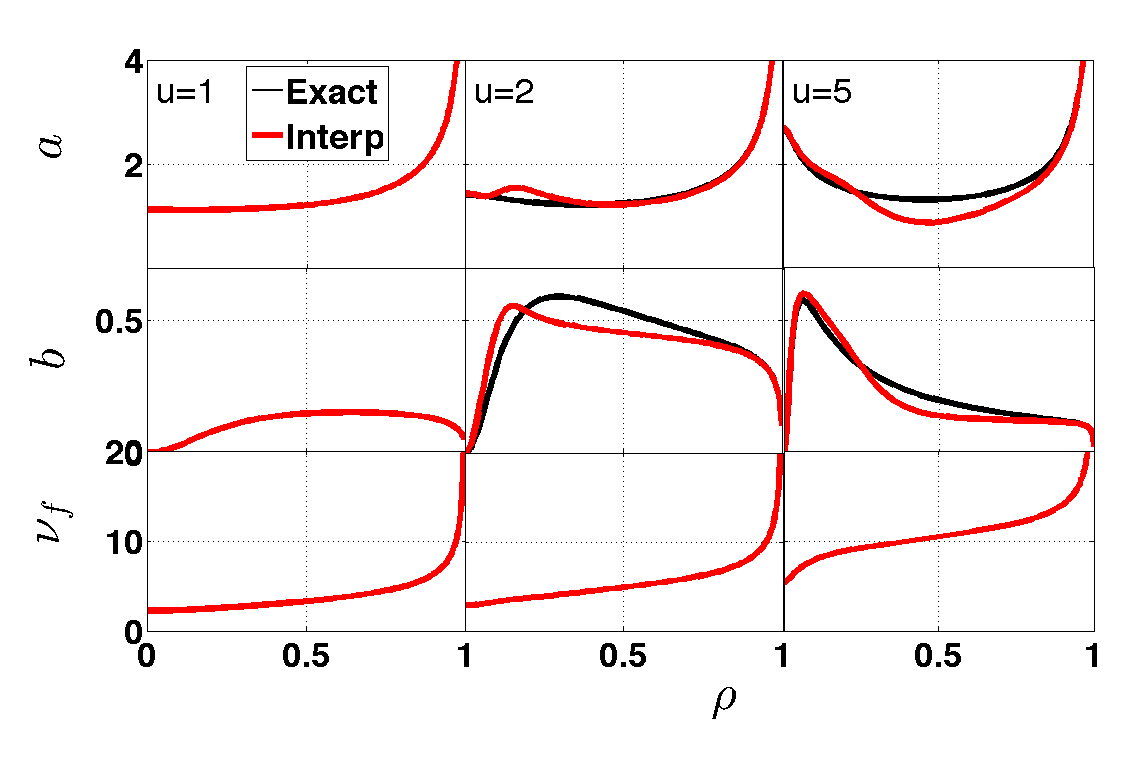}
\caption{Exact (solid black) and interpolated (red) Kernel parameters as a function(al) of density $\rho$.
The parameter $b$ for $u=5$ has been divided by 10 to fit in the same y-scale as in the other two panels.}
\label{fig:kernel_functional}
\end{figure}

\begin{figure}
\includegraphics[width=1.0\columnwidth]{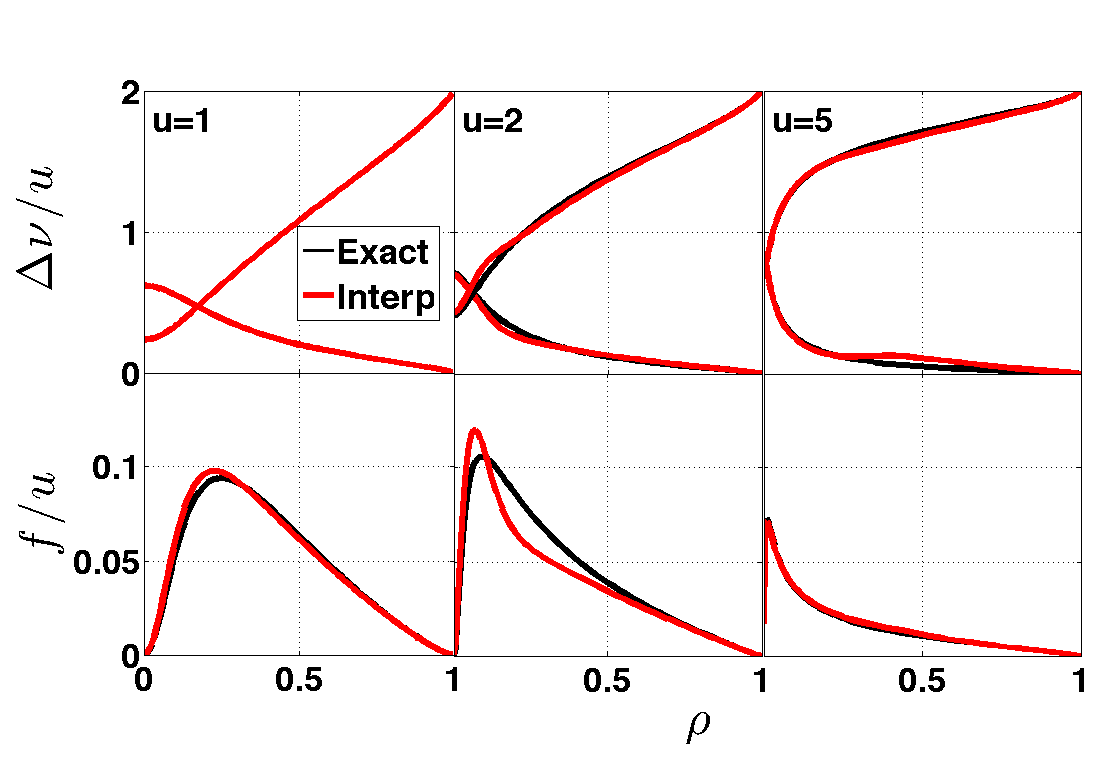}
\caption{Exact (solid black) and interpolated (red) frequency deviations and oscillator strength as a function of density $\rho$.}
\label{fig:freqs_functional}
\end{figure}

The final step is to write these interpolations as a function of $u$ and $\rho$
instead of the dependence on $x$ through $\bar z$.   This is accomplished again using the
results for the ${\cal F}$-functional from Appendix \ref{app:g0_review}.
We thus find  $z=x(\rho,u)/u=-1/u\times\partial f/\partial\rho$.
The values of $z(\rho,u)$ can be inserted into Eq. (\ref{eq:interpolation})
to deliver the kernel functional.  The kernel parameters as a function(al) of $\rho$
are plotted in Fig. \ref{fig:kernel_functional}. The frequency deviations and oscillator
strength as a function of $\rho$ are plotted in Fig. \ref{fig:freqs_functional}.

\sec{Discussion and outlook}

\ssec{Analogy to real diatomic molecules\label{sec:tp33}}
The asymmetric Hubbard dimer behaves similarly to
real diatomic molecules stretched to large bond-lengths 
when the latter are treated within a "minimal model", i.e. when only the KS HOMO and LUMO orbitals 
are considered. 
If the molecule is neutral, these two orbitals become energetically so close when approaching
the dissociation limit, that the minimal model captures the essential physics
since couplings to the many other orbitals in the molecule
are far smaller in comparison. 
In the Hubbard model, there are only ever two orbitals, so it makes a 
natural model for these stretched molecules. (Such a minimal model does
not capture van der Waal's interactions between the atoms, which result from fluctations
within each atom.)
The problem of laser-induced charge-transfer  dynamics has 
been studied in this way~\cite{FM14,FM14a,M17}.  Here we compare the kernel of the real molecule in this 
limit with that of the Hubbard model.

The ground state  of such a stretched neutral molecule has close to one electron on each atomic HOMO.  
Assuming then that the atomic orbitals are orthogonalized, we consider the MH 
limit of the Hubbard dimer, whose ground state is (see Appendix \ref{app:wavefunctions} 
for details)
\ben
\label{eq:tp4}
|\psi_0\rangle \approx  \frac{1}{\sqrt{2}}\,(|12\rangle+|21\rangle).
\een
(On the other hand, a stretched cationic diatomic molecule approaches
the CT limit of the Hubbard dimer, in the particular case where the
LUMO and HOMO of the molecule are on different atoms). 
In neutral molecules, the exact KS HOMO has the form of a  bonding orbital, straddling both
atoms with a density equal to the sum of the individual 
atomic HOMO densities, while the LUMO has approximately an antibonding form. This holds for both homo-atomic 
and hetero-atomic neutral molecules~\cite{GGGB00,M05c}.  
Their orbital energies become increasingly degenerate as the 
molecule is stretched, so the KS excitation energy becomes very small 
(exponentially small with the inter-atomic distance $R$).  This is 
consistent with the Hubbard dimer, where this excitation energy is equal to the hopping 
integral $2\,t$, which would also decay exponentially with $R$ (see again 
Appendix \ref{app:wavefunctions}). 
Strictly speaking, to model a heteronuclear neutral molecule with a Hubbard dimer, we should 
require different $U$-parameters on each site, with
$U_i = I_i - A_i$ (although Appendix \ref{app:U1U2} shows how to map such a dimer onto one with the same $U$ on each site). 
In any case, even with the same $U$ on each site, we 
capture the basic qualitative features of  excitations and the xc kernel of 
stretched molecules with the MH limit of the Hubbard dimer. 

For the molecule, we can write the kernel (in the minimal model, restoring
dimensional units)~\cite{M05c,M17}  
as $f\Hxc^{qq}(\omega) = f\Hxc^{qq}(\omega = 0) + f\Hxcd^{qq}(\omega)$. The adiabatic part
\ben
f\Hxc^{qq}(\omega = 0) = \frac{\omega_1\omega_2}{4\omega\s} - \frac{\omega\s}{4},
\een
where $\omega_1 = I_b - A_a - 1/R$, $\omega_2 = I_a - A_b - 1/R$ are the excitation frequencies for charge-transfer 
excitations from atom $b$ to atom $a$ and vice-versa,  and $\omega\s$  $\sim e^{-\alpha R}$ is the Kohn-Sham HOMO-LUMO gap. 
Comparing with the adiabatic Hubbard kernel in the MH limit, Eq. (\ref{eq:fst}),
\ben
f_{st} = \frac{\nu_1\,\nu_2}{2\,\nu_4} - \frac{\nu\s}{2\,W_s}
\een
we see the adiabatic part in both is proportional to the product of the exact excitation frequencies, and both blow up as in 
the limit ($u/x$ or $R \to \infty$). Comparing the dynamical part
\ben
f\Hxc^{qq,dyn}(\omega) = f\Hxc^{qq}(\omega) -f\Hxc(\omega = 0)  = \frac{\omega^2}{\omega\s}\left(\frac{\delta^2}{\omega^2 -\omega_1\omega_2} \right)\;.
\een
where $\delta = (\omega_1 - \omega_2)/2$, 
with that of the dimer, Eq. (\ref{eq:fdyn}), we observe both have a pole at the product of the two exact excitation frequencies, 
and both blow up in the limit. 
Thus the kernel in the case of a stretched diatomic molecule maps closely to the form of the kernel for 
the Hubbard dimer in the MH limit (Appendix~\ref{app:wavefunctions}). 

\ssec{Applications}

In this paper, we have thoroughly explored the linear response TDDFT of the
Hubbard dimer.   We have shown how the standard expansion of many-body
theory is not useful for understanding the competition between inhomogeneity
effects and correlation effects.  We find that strong correlation is better
characterized by an expansion in which the ratio $\dv/U$ is kept fixed rather than
$\dv$ itself.  It makes sense that inhomogeneity should be measured relative
to the interaction strength.  By expanding in powers of $1/u$ keeping that ratio
fixed, we find an accurate expansion for the strongly correlated limit.
Moreover, we can smoothly interpolate this expansion with the standard weakly-correlated
limit, and construct an explicit approximate XC kernel that works well
in both regimes, and does not fail badly in between.

How can this kernel be used?   Clearly, this kernel itself is constructed
within a lattice model, and so might be used as an approximation (or the
starting point of a more general approximation) to apply TDDFT to lattice
models.  There is substantial history of studies in this area\cite{AG02,B08,LU08,V08b,KSKVG10,T11,FFTAKR13,FT12,RP10,SDS13,FM14,FM14a,DSH18,KS18,KKPV13,KVOC11,MRHG14,TR14}.
Such
applications can be useful in studying systems too large to
be accessible by more direct quantum solvers, where the relative inexpensiveness
of DFT can be crucial.

A second way one could imagine this kernel being used is in a continuous
real-space calculation, e.g., a diatomic molecule, in which some choice
has been made that assigns some fraction of the electrons to each 
atom.  Then the kernel might be applied directly to these occupation numbers,
allowing double excitations to be included in TDDFT calculations of the 
system.  This might prove particularly effective when the bond is stretched,
so that electrons truly are localized on each site.

A third way the kernel might be used is simply as an illustration of the
effects of strong-correlation within linear-response TDDFT, to inspire
construction of frequency-dependent kernels that can be applied to realistic
systems.   Such kernels, when applied within a minimal basis model, should
capture the same effects shown here, as discussed in the previous section.

The range of validity of the kernel can be extended and tested by
solving larger or more complex systems like multi-orbital Hubbard dimers, because
some of these models are amenable to numerically exact solutions.
  
An important point in this work is also the literal existence of the kernel
itself.  We have given the explicit frequency-dependence of the dynamic
XC kernel that is exact for this Hamiltonian and two electrons.  Such
kernels do exist and reproduce the exact transition frequencies and
oscillator strength, including that of the double excitation, even when
it represents a charge transfer. 

The Hubbard dimer can be easily generalized to the asymmetric Anderson dimer
as discussed in Appendix \ref{app:U1U2}. So the results presented here can
be applied to this later model.

We have also proven or illustrated many smaller, related results, such 
as how to identify multiple excitations from single ones, the oscillator
strength sum-rule for this lattice model, the adiabatic connection formula
and the Kohn-Sham linear response for fractional occupations.

\acknowledgments
DC and JF wish to thank funding support from the Spanish Ministerio de Econom\'{\i}a y Competitividad via grant
FIS2012-34858.
NTM thanks the US National Science Foundation CHE-1566197 for support. 
KB acknowledges DOE grant number DE-FG02-08ER46496.

All the authors have benefitted either directly or indirectly from the accumulated
impact of Prof Gross's works.
Some also acknowledge about 45 accumulated years of friendship and learning at the feet of
Prof E.K.U. (Hardy) Gross, who taught us (almost) everything we know
about time-dependent density functional theory.  We hope that this small 
contribution, demonstrating the exactness of TDDFT in the simplest possible
case, might contribute to elucidating how the theory works
to skeptics in many-body theory or ab-initio
computational chemistry.  While this paper (and, indeed, much
of Prof Gross's work) might be regarded as FEPU (formally exact, practically
useless), the proof of the RG theorem\cite{RG84} was clearly anything but.

All authors have contributed to the article.

\bibliographystyle{apsrev}
\bibliography{master,hubbard,hubbardspan}
\appendix

\sec{Exact energies and weights}
\label{exact}
We use the following basis states to span the sub-space  labelled by $N=2$, $S^2=0$ and $S_z=0$:
\bea
\label{eq:basis}
|\varphi_a\rangle&=&\frac{|12\rangle+|21\rangle}{\sqrt{2}}=(1,\,0,\,0)^\dagger\nonumber\\
|\varphi_b\rangle&=&|11\rangle=(0,\,1,\,0)^\dagger\\
|\varphi_c\rangle&=&|22\rangle=(0,\,0,\,1)^\dagger\nonumber
\eea
\ssec{Many-body states}
\label{app:eigenvaules}
The three singlet eigen-energies of the Hubbard dimer within the sub-space
are ($i=0,1,2$):
\bea
e_i&=&\frac{2}{3}\left(u + \sqrt{3+3\,x^2+u^2}\,\cos\left(\theta+\frac{2\,\pi}{3}\,(i+1)\right)\right),\nonumber\\
\theta&=&\frac{1}{3}\,cos^{-1}\left[\frac{9\,x^2-9/2-u^2}{(3+3\,x^2+u^2)^{3/2}}\,u\right]
\label{eq:theta}
\eea
and $cos^{-1}$ denotes the principal value of the complex arccos function.
Next, the eigenstates are,
\ben
|\Psi_i\rangle=(\alpha_i,\,\beta_i^+,\,\beta_i^-)^\dagger,
\nonumber
\een
\ben
\alpha_i=\frac{e_i-u}{e_i\,r_i},~~
\beta^\pm_i=\frac{u-e_i\pm x}{\sqrt{2}\,r_i},
\een
\ben
r_i=r(e_i)=\sqrt{x^2+(e_i-u)^2\,\left(1+1/e_i^2\right)}.\nonumber
\een
Notice that normalization implies
$|\alpha_i|^2+|\beta_i^+|^2+|\beta_i^-|^2=1$, and that the density for each of the three states 
is $\rho_i=\Delta n_i/2=(\beta_i^+)^2-(\beta_i^-)^2$. We denote the ground-state density 
$\Delta n=\Delta n_0$ (or $\rho=\rho_0$) and
the transition frequencies  as 
\bea
\label{eq:tc1}
\nu_1&=&e_1-e_0=2\,\sqrt{1+x^2+u^2/3}\,\sin\,\theta,\\
\nu_2&=&e_2-e_0=2\,\sqrt{1+x^2+u^2/3}\,\sin(\theta+\pi/3).\nonumber
\eea
The weights are given by
\ben
\label{eq:tc2}
{\sqrt{W_{1,2}}} = 
|\langle \psi_{0}|\Delta\hat{n}|\psi_{1,2}\rangle|=\frac{4\,x\,e_{2,1}}{r(e_0)\,r(e_{1,2})},
\een
while
\ben
\label{eq:tc3}
\nu_3=\nu_1\,W_1+\nu_2\,W_2
=-\frac{8\,(e_0-u)^2}{e_0\,r_0^2}.
\een
Eqs. (\ref{eq:tc1}) and (\ref{eq:tc2}) are used in Eq. (\ref{eq:S_def}), and Eq. (\ref{eq:tc3}) is used in Eq. (\ref{eq:sumrule}) of the main text.

\ssec{Kohn-Sham states}
The spin-independent dimensionless Hamiltonian written in the single-particle $\{|1\rangle,|2\rangle\}$ basis is
\bea
\hat{h}_s&=&
\left(\begin{array}{cc}\frac{\bar{v}_s}{2\,t}-\frac{\Delta v_s}{4t}&-1/2\\-1/2&
\frac{\bar{v}_s}{2\,t}+\frac{\Delta v_s}{4t}\end{array}\right),\,
\eea
where the KS potentials are 
\bea
\bar{v}_s&=&\bar{v}+\bar{v}_{Hxc}=\bar{v}_{Hxc},\\
\frac{\Delta v_s}{2t}&=& \frac{\Delta v}{2t}+\frac{\Delta v_{Hxc}}{2t}=x_s=x+x_{Hxc}.\nonumber
\eea
It is useful to define the auxiliary variables  $r_s=\sqrt{x_s^2+1}$,  $\bar{x}_s=x_s/r_s$. Then,  
the eigenvalues and normalized eigenfunctions are given by
\ben
e_s^\pm=\bar{v}_s/(2\,t)\pm r_s/2,\,\,\,|\phi^\pm\rangle=c_s^\pm\,|1\rangle\mp c_s^\mp\,|2\rangle,
\een
where $c_s^\pm=\sqrt{\left(1\mp \bar{x}_s\right)/2}$.
The ground-state density is
\ben
\label{eq:tc4}
\rho=\langle\phi_0|\frac{\Delta\hat{n}}{2}|\phi_0\rangle=\bar{x}_s,\,\,\,x_s=\frac{\rho}{\sqrt{1-\rho^2}}.
\een
The singlet KS 2-particle states can be found from Slater determinants of
the KS single-particle states:
 \bea
 |\Phi_0\rangle&=&\left(\sqrt{2}\,c_s^+\,c_s^-,\,(c_s^-)^2,\,(c_s^+)^2\right)^\dagger,\nonumber\\
 |\Phi_1\rangle &=&\left((c_s^+)^2-(c_s^-)^2,\,\sqrt{2}\,c_s^+\,c_s^-,\,-\sqrt{2}\,c_s^+
 \,c_s^-\right)^\dagger,\\
  |\Phi_2\rangle&=&(\sqrt{2}\,c_s^+\,c_s^-,-\,(c_s^+)^2,\,-(c_s^-)^2)^\dagger.\nonumber\\
 \eea
The KS transition frequencies are:
\ben
\label{eq:tc5}
\nu_{s} = r\s,~~~~\nu_{d} = 2 \,r\s,
\een
where $\nu_{d}$ is the KS double, trivially twice the single, $\nu_{s}$. The weights are
\ben
\label{eq:tc6}
{\sqrt{W_{s}}} = 
|\langle \Phi_0|\Delta\hat{n}|\Phi_1\rangle|=\frac{\sqrt{2}}{r_s},
\een
while $W_{d}$ vanishes entirely. Eqs. (\ref{eq:tc4}), (\ref{eq:tc5}) and (\ref{eq:tc6}) are used in Eq. (\ref{eq:tp1}) of the main text.

\sec{Proofs and generalizations}

\ssec{Oscillator strength sum rule}
\label{app:sumrule}
The sum rule for the density-density response operator can be obtained from: 
\ben
\langle\Psi_0|\,[\hat{\rho},[\hat{\rho},\frac{\hat{H}}{2\,t}]]\,|\Psi_0\rangle=-2\,
\sum_{m\neq 0}\,\nu_m\,|\langle\Psi_0|\,\hat{\rho}\,|\Psi_m\rangle|^2.
\een
Some little algebra shows that the commutators can be written as $\hat{T}/(2\,t)$, yielding
\ben
\label{eq:sumrule0}
\sum_{m\neq 0}\,\nu_m\,|\langle\Psi_0|\,\hat{\rho}\,|\Psi_m\rangle|^2=-\frac{1}{2}\,\langle\Psi_0|\,\frac{\hat{T}}{2\,t}\,|\Psi_0\rangle.
\een
This result is general and valid for the Hubbard dimer irrespective of the number of electrons. The relation 
between the kinetic energy and the weights of the density-density linear response in the Hubbard model has 
been already established in the literature in the past (see e.g. [\onlinecite{M77,BGR87}]). In these references
it is emphasized that the sum rule for this model is not providing the full story because the Hamiltonian contains 
only a single state per site and thus allows only intraband transitions. The complete $f$-sum rule includes all 
allowed interband transitions and does not depend on the electron-electron interaction unlike the case in the 
Hubbard model \cite{BGR87}.
Eq. (\ref{eq:sumrule0}) reads explicitly for $N=2$
\ben
\label{eq:sumrule2}
\nu_3=\nu_1\,W_1+\nu_2\,W_2=-\frac{8\,(e_0-u)^2}{e_0\,r_0^2},
\een
where the right-hand side of the equation is a function of $x$ and $u$. Eq. (\ref{eq:sumrule2}) is used in Eq. (\ref{eq:sumrule}) of the main text.

\ssec{Fluctuation-dissipation theorem}
\label{app:FD}
We start by rewriting the Hubbard interaction term in terms of $N$ and $\dn$,
\bea
\hat{V}_{ee}=\frac{U}{4}\,(\hat{N}^2+\Delta\hat{n}^2)-\frac{U\,\hat{N}}{2}
\eea
where we have used the fact that $\hat{n}_{i\sigma}^2=\hat{n}_{i\sigma}$ for fermion operators.
Using this definition we can write the Hamiltonian 
\bea
\hat{{\cal H}}^\lambda&=&\hat{V}^\lambda+\hat{T}+\lambda\,\hat{V}_{ee}\\
&=&-\frac{\Delta v^\lambda\,\Delta\hat{n}}{2}+\hat{T}+\frac{\lambda\,U}{4}\,\left(\hat{N}^2+\Delta\hat{n}^2\right)-\frac{\lambda\,U\,\hat{N}}{2}.\nonumber
\eea
By integrating the Hellmann-Feynman equation between $\lambda=0$ and $\lambda=1$ we obtain the following expression for the ground-state
energy
\bea
\label{eq:E0AC}
E_0&=&-\frac{\Delta n\,\Delta v}{2}+T_s\\
&&+\frac{U}{4}\,\int_0^1\,d\lambda\,\langle\Psi_0^\lambda|\hat{N}^2+\Delta\hat{n}^2|\Psi_0^\lambda\rangle-\frac{U\,N}{2}.\nonumber
\eea
By comparing this expression for $E_0$ with the definition of the total energy, we extract
\ben
\label{eq:ExcAC}
E_{xc}=\frac{U}{4}\,\int_0^1\,d\lambda\,\langle\Psi_0^\lambda|\hat{N}^2+\Delta\hat{n}^2|\Psi_0^\lambda\rangle-U_H-\frac{U\,N}{2}.
\een
The first term in the integrand in Eq. (\ref{eq:E0AC}) is just $N^2$, while from  Eq. (\ref{eq:chi_dimer}) we find
\ben
\sum_{m\neq 0}\,|\langle\Psi_0^\lambda|\Delta\hat{n}|\Psi_m^\lambda\rangle|^2=-\frac{1}{\pi}\int_0^\infty\,d\omega\,Im\,\chi^\lambda(\omega).
\een
Inserting this into Eq. (\ref{eq:ExcAC}) we finally have,
\ben
\label{eq:tc11}
E_{xc}=-\frac{U}{4\,\pi}\,\int_0^1\,d\lambda\,\int_0^\infty\,d\omega\,Im\,\chi^\lambda(\omega)-\frac{U\,N}{2}
\een
where we have made use of the expression of the Hartree energy,
\ben
U_H=\frac{U}{4}\,\left(N^2+\dn^2\right).
\een
We also see that
\ben
E_x=-\frac{U}{4\,\pi}\,\int_0^\infty\,d\omega\,Im\,\chi^{\lambda=0}(\omega)-\frac{U\,N}{2},
\een
where we have made use of the expression of the exchange energy for integer occupations $N =1,\, 2$,
\ben
E_x=-\frac{U}{4\,N}\,\left(N^2+\dn^2\right).
\een
This finally yields,
\ben
E_c=-\frac{U}{4\,\pi}\,\int_0^1\,d\lambda\,\int_0^\infty\,d\omega\,Im\,\left(\,\chi^\lambda(\omega)-\chi^{\lambda=0}(\omega)\right).
\een
Eq. (\ref{eq:tc11}) is used to define Eq. (\ref{eq:tp11}) of the main text.

\ssec{Generalization to  $U_i$}
\label{app:U1U2}
It is easy to show that any result obtained for the Hubbard dimer can be easily translated to a dimer with 
different Coulomb energies $U_1$ and $U_2$ by simply re-writing the on-site potential and Coulomb terms. For example
for $N=2,S^z=0$ we can use the relationships
\ben
\label{eq:tc66}
\dv=\dv'+\frac{U_2-U_1}{2},\,\,\,\,U=\frac{U_2+U_1}{2},
\een
to write
\begin{equation}
\hat{H}=\left(\begin{array}{ccc}
0&-\sqrt{2}\,t&-\sqrt{2}\,t \\ -\sqrt{2}\,t &-\dv'+U_1 &  0 \\ -\sqrt{2}\,t&0&\dv'+U_2
\end{array}\right).
\end{equation}
Similar transformations can be defined for $N=1,3$.
A corollary is that the solution of the Anderson dimer can be obtained from the solution of its equivalent asymmetric Hubbard dimer.
Eq. (\ref{eq:tc66}) can be inserted in Eq. (\ref{eq:tp2}) of the main text.

\ssec{Fractional particle number}
\label{app:fractional}
By inverting the relation
\ben
\Delta n[\Delta v_s,{\cal N}]=(1-w)\,\Delta n[\Delta v_s,N]+w\,\Delta n[\Delta v_s,N+1]
\een
where ${\cal N}=N+w$, and defining $\tilde {\cal N}={\cal N}$ for ${\cal N}\leq 2$, $\tilde {\cal N}=4-{\cal N}$ for 
${\cal N}\geq 2$, we find 
\bea
\frac{\Delta v_s[\dn,{\cal N}]}{2\,t}&=&
\frac{\Delta n}{\sqrt{\tilde{\cal N}^2-\Delta n^2}},\nonumber\\
\label{eq:rsdn}
r_s[\dn,{\cal N}]&=&
\frac{\tilde{\cal N}}{\sqrt{\tilde{\cal N}^2-\Delta n^2}},\\
c_s^\pm&=&
\frac{1}{\sqrt{2}}\,\left(1\mp\frac{\dn}{\tilde{\cal N}}\right)^{1/2}.\nonumber
\eea
The above expressions yield
\ben
\label{eq:chis}
\chi_{s}(\nu)=
\frac{4\,\sqrt{\tilde{\cal N}^2-\Delta n^2}}{\tilde{\cal N}\,\left(\nu^2-\frac{\tilde{\cal N}^2}{\tilde{\cal N}^2-\Delta n^2}\right)},
\een
that  indicates that we can generalize the response function to arbitrary fractional ${\cal N}$.
Eq. (\ref{eq:chis}) is used to define the exact expressions of the coefficients in Eq. (\ref{eq:tp1}) of the main text.

\sec{Expansions and limits}
\label{app:expansions}
\ssec{Symmetric limit}
\label{sym}
The energies can be written in terms of $r_u=\sqrt{4+u^2}$ as
\ben
e_{0,2}=\frac{1}{2}\,(u\mp r_u),\,\,\,e_1=u.~~~{\rm (sym)}
\een
The linear response frequencies and weights are
\bea
\label{eq:tp10}
\nu_1&=&\frac{1}{2}\,(u+r_u),\,\,\,\,\,\,\,\nu_2=r_u,~~~{\rm (sym)}\\
W_1&=&2\,\left(1-\frac{u}{r_u}\right),\,\,\,W_2=0.\nonumber
\eea
The weight of the second excitation is identically zero. The linear response parameters described in the main text are
\ben
a=\frac{r_u}{8},\,\,\,\,\,\,b=0,\,\,\,\,\,c=\frac{r_u}{8}\,\frac{r_u+u}{r_u-u},\,\,\,\,\,\nu_f=r_u.~~~{\rm (sym)}
\een
Finally, the overlap between the exact and KS ground state wavefunctions is
\ben
\langle\Psi_0|\Phi_0\rangle=\frac{2-u+r_u}{\sqrt{2\,\left[(u-r_u)^2+4\right]}}.~~~{\rm (sym)}
\een
Eqs. (\ref{eq:tp10}) is used in the discussions after Eq. (\ref{eq:S_def}) and Fig. \ref{freq}.
\ssec{Ground-state density functional}
\label{gs}
\label{app:g0_review}
The ${\cal F}-$functional of the Hubbard dimer looks like
\bea
{\cal F}(\rho, u)&=&\frac{F(\rho,u)}{2\,t}=\underset{\Psi}{min}\,\langle\Psi\,|\,\frac{\hat{T}}{2t}+\frac{\hat{V}_{ee}}{2t}\,|\,\Psi\rangle\nonumber\\
&=&\underset{g}{min}\, (-g + h(g,\rho,u)),
\label{eq:geq}\\
h(g,\rho,u)&=&u \,\frac{g^2\,\left(1-\sqrt{1-g^2-\rho^2}\right)+2\,\rho^2}{2\,(g^2+\rho^2)}.\nonumber
\eea
Solving  for $g$ in $\partial h/\partial g =1$ yields a tenth-order equation, that after some tuning can be reduced to
the following sixth-order equation
\bea
\label{eq:pol}
p_2(g)\,u^2+p_1(g)\,u+p_0(g)=0,
\eea
where
\bea
p_0&=&(g^2+\rho^2)^2\,(g^2+\rho^2-1),\nonumber\\
p_1&=&2\,\rho^2\,g\,(g^2+\rho^2-1),\\
p_2&=&g^2\,((g^2/2+\rho^2)^2-\rho^2).\nonumber
\eea
The resulting $g_0$, when introduced in
Eq.  (\ref{eq:geq}) delivers the ${\cal F}-$functional. This is substituted in the equation
$\partial {\cal F}/\partial \rho=-x$   to find $z(\rho)=x/u$. 

We bring back now the ansatz developed in Ref. \cite{CFSB15}, that provides an excellent approximation
for the reduced potential $z(\rho)$. This is
\bea
g_0^{app}(\rho)&=&\sqrt{\frac{(1-\rho)\,(1+\rho\,(1+(1+\rho)^3\,a_1\,u))}{1+(1+\rho)^3\,a_2\,u}},\nonumber\\
a_i&=&a_{i1}+a_{i2}\,u,\\
a_{21}&=&\frac{\sqrt{(1-\rho)\,\rho/2}}{2},\,\,a_{11}=(1+\rho^{-1})\,a_{21},\nonumber\\
a_{12}&=&\frac{1-\rho}{2},\,a_{22}=\frac{a_{12}}{2}\nonumber.
\eea
We show in Fig. \ref{fig:zdn_cfsb} in the main text that the potential $z(\rho)$ obtained this way provides a very accurate 
fit to the exact reduced potential.

\sssec{Weakly correlated functional expansion}
\label{app:rhoWC}
We expand the parameter $g$  using the weak coupling expansion $g=\sum_n a_n\,u^n$, 
and then apply the constraint $\partial h/\partial g=1$ to find the $a_n$ coefficients for $g_0$. We find
\bea
\label{eq:tc10}
g_0&=&\bar{\rho}\,\left(1-\frac{\bar{\rho}^2\,u^2}{8}+\frac{\rho^2\,\bar{\rho}^{5/2}\,u^3}{4}\right),\nonumber\\
{\cal F}&=&-\bar{\rho}\,\left(1-\frac{(1+\rho^2)\,u}{2}+\frac{\bar{\rho}^2\,u^2}{8}-\frac{\rho^2\,\bar{\rho}^{5/2}\,u^3}{8}\right),\\
\left|z^{WC}\right|&=&\frac{\rho}{\bar{\rho}\,u}\,\left(1+\rho\,u+\frac{5\,\bar{\rho}^2\,u^2}{8}+\frac{(1-4\,\rho^2)\,\bar{\rho}^{5/2}\,u^3}{4}\right),\nonumber
\eea
where $\bar{\rho}=\sqrt{1-\rho^2}$.
This procedure delivers an accurate estimate of $z(\rho)$ for $u \leq 1-2$. We find that adding
higher orders than $u$ spoils the estimate.
Eq. (\ref{eq:tc10}) is used in Eq. (\ref{eq:tp100}) of the main text.

\sssec{Strongly correlated functional expansion}
The large-$u$ expansion can be found from eq. (\ref{eq:pol}). We expand $g=\sum_n b_n\,u^{-n}$ and find
\bea
g_0&=&\tilde{\rho}\,\left(1+\sqrt{\frac{1-\rho}{2\,\rho}}\,\frac{1}{2\,u}\right.\nonumber\\
&&\left.+\frac{3\,(1-3\,\rho)}{16\,\rho\,u^2}+\frac{1-8\,\rho+11\,\rho^2}{8\,\rho\,\tilde{\rho}\,u^3}\right),\nonumber\\
\frac{\cal F}{u}&=&\rho\,\left(1-\frac{\tilde{\rho}}{\rho\,u}-\frac{1-\rho}{4\,\rho\,u^2}-\frac{(1-3\,\rho)\,\tilde{\rho}}{16\,\rho^2\,u^3}\right),\\
\left|z^{SC}\right|&=&1-\frac{1-2\,\rho}{\tilde{\rho}\,u}+\frac{1}{4\,u^2}+\frac{1+3\,\rho-6\,\rho^2}{16\,\tilde{\rho}\,\rho\,u^3},\nonumber
\eea
with $\tilde{\rho}=\sqrt{2\,\rho\,(1-\rho)}$.
This procedure provides an accurate estimate of $z(\rho)$ for sufficiently large $u$, except near $\rho= 0$. We have
found that including higher orders in the expansion also spoils how $z^{SC}$ fits $z$. This appendix is not used in the main text, 
but is included for completeness.

\ssec{Many-body expansion}
\label{app:Useries}
The Taylor series expansion in powers of  $u$ for fixed $x$ can be found by straightforward perturbation theory.
A simpler route however consists of expanding $\theta$ in Eq. (\ref{eq:theta}) in powers of $u$.
We find to the order given:
\begin{eqnarray}
e_{0,2}&=&\mp p_0\,\left( 1\mp \left(\half+x^2\right)\,\tilde{u}+ \frac{1/4+x^2}{2}\,\tilde{u}^2\right),\nonumber\\
e_1&=&p_0\,\tilde{u}\,\left(1+x^4\,\tilde{u}^2\right),
\end{eqnarray}
where $p_0=\sqrt{1+x^2}$ and $\tilde{u}=u/p_0^3$.
The frequencies are
\ben
\label{eq:omegaSwc}
\frac{\nu_j}{p_0}=j\left(1+
\frac{1+4\,x^2}{8}\,\tilde{u}^2\right)+\delta_{j1}
\frac{1-2\,x^2}{2}\,\tilde{u},
\een
while the weights are
\bea
\label{eq:Swc}
W_1&=&\frac{1}{p_0^2}\,\left(2+(4\,x^2-1)\,\tilde{u}+2\,x^2\,(3\,x^2-4)\,\tilde{u}^2\right),\nonumber\\
W_2&=&\frac{x^2}{p_0^2}\,\tilde{u}^2\,\left(1+\left(2\,x^4-4\,x^2-\frac{1}{4}\right)\,\tilde{u}^2\right),
\eea
and the oscillator strength is
\ben
f=x^2\,\tilde{u}^2.
\een
The KS values are
\bea
\frac{\nu_s}{p_0}&=&1-x^2\,\tilde{u}+\frac{7\,x^2}{8}\,\tilde{u}^2,\nonumber\\
W_s&=&\frac{2}{p_0^2}\,\left(1+2\,x^2\,\tilde{u}+x^2\,\left(3\,x^2-\frac{7}{4}\right)\,\tilde{u}^2\right).
\eea
The kernel parameters are, to the order given,
\bea
\label{eq:kernelpu2}
a&=&\frac{p_0}{4}\,\left(1-x^2\,\tilde{u}+\left(x^2+\frac{1}{8}\right)\,\tilde{u}^2\right),\nonumber\\
b&=&\frac{9\,x^2\,p_0^3\,\tilde{u}^2}{16}\,\left(1-\frac{2}{3}\,p_0\,\tilde{u}\right),\\
\nu_f&=&2\,p_0\,\left(1+\frac{p_0^2\,\tilde{u}^2}{8}\right).\nonumber
\eea
Eq. (\ref{eq:kernelpu2}) is used to define Eq. ({\ref{eq:kernel2}) of the main text.

\ssec{Expansion for fixed interaction-asymmetry ratio}

We find that  Eq. (\ref{eq:theta})
can be written as the following cubic equation for the variable $\cos\theta$: 
\bea
\cos 3\,\theta =\,\cos\theta\,\left(4\,\cos^2\theta-3\right)&=&
\frac{9\,z^2-1-9/(2\,u^2)}{(z_b+3/u^2)^{3/2}}\nonumber\\
\eea
where $z_b=1+3\,z^2$.
The zeroth order of the above equation in a $1/u$ expansion looks
hardly solvable for $\cos\theta$:
\bea
\label{eq:cubic}
\cos\theta\,\left(4\,\cos^2\theta-3\right)=
\frac{9\,z^2-1}{z_b^{3/2}}~~(\rm{u\rightarrow\infty}).
\eea
However, we note that the second excited
state can be written in this limit as
\bea
e_2=u\,(1+z)=\frac{2}{3}\,u\,(1+z_b^{1/2}\,\cos\theta)
\eea
so that we find
\bea
2\,z_b^{1/2}\,\cos\theta(z)&=&1+3\,z\\
2\,z_b^{1/2}\,\sin\theta(z)&=&\pm\sqrt{3}\,(1-z).\nonumber
\eea
It is easy to check that this result for $\cos\theta$ solves the cubic
equation (\ref{eq:cubic}).  
Choosing the plus or minus signs for the $\sin$ function yield $e_0=0$ or $u\,(1-z)$ hence rendering the MH or CT regimes,
respectively. 
We can expand now the full $\theta$ function
in powers of $1/u^2$, and retrieve easily the results found below using the perturbation theory. These results are used to find Eq. (\ref{eq:kernel2}) in
Section \ref{sec:tc11} and Eqs. (\ref{eq:MH}) in Section \ref{sec:tc12} and Eq. (\ref{eq:kernel4}) in Section \ref{sec:tc13}.

\label{app:tSeries}
\sssec{Perturbative expansion}
The basis states $|\varphi_{a,b,c}\rangle$ defined in Eq. (\ref{eq:basis}) become the eigenstates for $u=\infty$, and are the starting point of the perturbative expansion. The ground state is 
$|\Psi_0\rangle=|\varphi_a\rangle$ if  $z<1$ and  
$|\Psi_0\rangle=|\varphi_b\rangle$ if  $z>1$. There
is therefore a change of limits at $z=1$ that demands a different expansion for the MH and CT regimes. 
The dimensionless perturbed energies to third order in $1/u$ are
\begin{eqnarray}
e_a&=&-\frac{1}{z_a\,u}+\frac{16\,(z^2+1)}{(z_a\,u)^3},\\
e_{b,c}&=&u\,(1\mp z)+\frac{1}{2\,(1\mp z)\,u}-\frac{z\pm 1}{8\,z\,((1\mp z)\,u)^3}.\nonumber
\end{eqnarray}
where $z_a=1-z^2$, and the corresponding perturbed states to up to order $1/u^4$ are:
\bea
\alpha_i&=&g_{\alpha,i}\,\sum_k\,f_{\alpha,i}^{(k)}\,u^{-k},\nonumber\\
\beta_i^{\pm}&=&g_{\beta,i,\pm}\,\sum_k\,f_{\beta,i,\pm}^{(k)}\,u^{-k},\nonumber\\
g_{\alpha,a}&=&g_{\beta,b,+}=g_{\beta,c,-}=1,\nonumber\\
g_{\alpha,b}(z)&=&-\frac{1}{\sqrt{2}\,(1- z)},\,\,g_{\alpha,c}(z)=g_{\alpha,b}(-z),\nonumber\\
g_{\beta,a,\pm}&=&\mp\frac{1}{\sqrt{2}\,(z\mp 1)},\nonumber\\
g_{\beta,b,-}(z)&=&-\frac{1}{4\,z\,(1-z)},\,\,g_{\beta,c,+}(z)=g_{\beta,b,-}(-z),\nonumber\\
f_{\alpha,a}^{(0)}&=&f_{\alpha,b}^{(1)}=f_{\beta,a,\pm}^{(1)}=f_{\beta,b,+}^{(0)}=f_{\beta,b,-}^{(2)}=\nonumber\\
&=&f_{\beta,c,+}^{(2)}=f_{\beta,c,-}^{(0)}=1,\nonumber\\
f_{\alpha,a}^{(2)}&=&-\frac{z^2+1}{2\,z_a^2},\\
f_{\alpha,a}^{(4)}&=&\frac{3\,z^4+30\,z^2+11}{8\,z_a^4},\nonumber\\
f_{\alpha,b}^{(3)}(z)&=&-\frac{2\,z+ 1}{4\,z\,(1- z)^2},\,\,f_{\alpha,c}^{(3)}(z)=f_{\alpha,c}^{(3)}(-z),\nonumber\\
f_{\beta,a,\pm}^{(3)}&=&-\frac{z^2\pm 2\,z+3}{2\,z_a^2},\nonumber\\
f_{\beta,b,+}^{(2)}(z)&=&-\frac{1}{4\,(1-z)^2},\,\,f_{\beta,c,-}^{(2)}(z)=f_{\beta,b,+}^{(2)}(-z),\nonumber\\
f_{\beta,b,+}^{(4)}(z)&=&\frac{6\,z^2+6\,z-1}{32\,z^2\,(1-z)^4},\,\,f_{\beta,c,-}^{(4)}(z)=f_{\beta,b,+}^{(4)}(-z),\nonumber\\
f_{\beta,b,-}^{(4)}(z)&=&-\frac{3}{4\,(1-z)^2},\,\,f_{\beta,c,+}^{(4)}(z)=f_{\beta,b,-}^{(4)}(-z).\nonumber
\eea

\sssec{Mott-Hubbard regime $U > \dv$}
The ordering of states in the MH regime is $0=a$, $1=b$, $2=c$.
Then, the excitation energies  are:
\bea
\frac{\nu_{1,2}\MHfour}{u}&=&
1\mp z\pm \frac{z\pm 3}{2\,z_a\,u^2}\mp\frac{(1\pm z)^4\pm 8\,z\,(z^2+1)}{8\,z\,z_a^3\,u^4},\nonumber\\
\eea
while the weights and oscillator strength are
\bea
\label{eq:tc14}
W_{1,2}\MHfour&=&\frac{2}{(1\mp z)^2\,u^2}-\frac{2\,z^3\pm 7\,z^2+8\,z\mp 1}{z\,(1-z)^4\,(z+1)^2\,u^4}\\&&
\pm\frac{2\,(z^5\pm 8\,z^4+23\,z^3\pm 20\,z^2+14\,z\mp 2)}{z\,(1-z)^6\,(z+1)^4\,u^6},\nonumber\\
f\MHfour&=&\frac{1-z}{2}+\frac{3\,z^2-1}{4\,z\,z_a\,u^2}-\frac{3\,z^4-22\,z^2+3}{16\,z\,z_a^3\,u^4}.\nonumber
\eea
The density $\rho=(\beta_0^+)^2-(\beta_0^-)^2$ is given to second order in $t$ by
\bea
\rho&=&\frac{2\,z}{(z_a\,u)^2}\,\left(1-\frac{2\,z\,(2+z^2)}{(z_a\,u)^4}\right).
\label{eq:dnMHt0}
\eea
This formula fits very well the exact $\rho$, although a slight improvement can be gained by using
\bea
\rho&=&\frac{2\,z}{z^2+u^2\,z_a^2}.
\label{eq:dnMH1t0b}
\eea
The kernel parameters are
\bea
\label{eq:tc12}
a\MH6&=&\frac{z_a\,u}{8}\,\left(1+\sum_{p=1}^3\,\frac{f_a^{(p)}(z)}{(z_a\,u)^{2\,p}}\right),\nonumber\\
b\MH6&=&\frac{z_a^2\,z^2\,u^3}{2\,z_b}\,\left(1+\sum_{p=1}^3\,\frac{f_b^{(p)}(z)}{(z_a^2\,z_b\,u^2)^p}\right),\nonumber\\
\nu_f\MH6&=&z_b^{1/2}\,u\,\left(1+\sum_{p=1}^3\,\frac{f_\nu^{(p)}(z)}{(z_a\,z_b\,u^2)^p}\right)\nonumber\\
f_a^{(1)}&=&2\,(1+z^2),\nonumber\\
f_a^{(2,3)}&=&-2-7\,z^2+z^4,\,\,2\,(1+6\,z^2+z^4),\nonumber\\
f_b^{(1)}&=&7\,z^4+18\,z^2-1,\\
f_b^{(2)}&=&\frac{37\,z^8-152\,z^6+202\,z^4-352\,z^2+9}{4},\nonumber\\
f_b^{(3)}&=&-\frac{13+1063\,z^4+219\,z^8+49\,z^{12}}{2}\nonumber\\
&&+383\,z^2+1242\,z^6+71\,z^{10},\nonumber\\
f_\nu^{(1)}&=&2-z^2,\nonumber\\
f_\nu^{(2)}&=&-\frac{4-3\,z^2+19\,z^4-4\,z^6}{2},\nonumber\\
f_\nu^{(3)}&=&\frac{8-4\,z^2+137\,z^4+70\,z^6+49\,z^8-4\,z^{10}}{2}\nonumber
\eea
where $z_a=(1-z^2)$, $z_b=3\,z^2+1$.
Eq. (\ref{eq:tc12}) is used in Eq. (\ref{eq:MH}) of the main text. Eqs. (\ref{eq:tc14}) and  (\ref{eq:tc12}) are used to plot Fig. \ref{fig:comparison_largeu_z} of the main text.

\sssec{Charge Transfer regime $U < \dv$}
The ordering of states in the CT regime is $0=b$, $1=a$, $2=c$.
Then, the excitation energies, weights and strengths are:
\bea
\label{eq:tc13}
\nu_1&=&u\,(z-1)+\frac{z+3}{2\,\bar{z}_a\,u}\nonumber\\
&&-\frac{z^4+12\,z^3+6\,z^2+12\,z+1}{8\,z\,\bar{z}_a^3\,u^3},\nonumber\\
\nu_2&=&
2\,z\,u+\frac{z}{\bar{z}_a\,u}-\frac{z^4+6\,z^2+1}{4\,z\,\bar{z}_a^3\,u^3},\nonumber\\
W_1&=&\frac{1}{(z-1)^2\,u^2}\,\left(2-\frac{2\,z^3+7\,z^2+8\,z-1}{z\,\bar{z}_a^2\,u^2}\right.\\&&
\left.+\frac{2\,(z^5+8\,z^4+23\,z^3+20\,z^2+14\,z-2)}{z\,\bar{z}_a^4\,u^4}\right),\nonumber\\
W_2&=&\frac{1}{z^2\,\bar{z}_a^2\,u^4}\,\left(1
-\frac{2\,(2\,z^2+1)}{\bar{z}_a^2\,u^2}\right.\nonumber\\&&\left.
+\frac{40\,z^6+95\,z^4+26\,z^2-1}{4\,z^2\,\bar{z}_a^4\,u^4}\right),\nonumber\\
f&=&\frac{1}{z\,\bar{z}_a\,(z+1)\,u^2}
-\frac{6\,z^3-3\,z^2+2\,z-1}{2\,z^2\,\bar{z}_a^3\,(z+1)\,u^4}.\nonumber
\eea
Here $\bar{z}_a=z^2-1$. The density is given to second order in $t$ by
\ben
\dn=2-\frac{1}{(z-1)^2\,u^2}+\frac{3\,z^2+4\,z-1}{4\,z^2\,(z-1)^4\,u^4},
\label{eq:dnCTt0bb}
\een
although the following expression fits the exact $\dn$ better:
\ben
\dn=\frac{4\,u^2\,(1-z)^2}{2\,u^2\,(1-z)^2+1}.
\label{eq:dnCT1t0b1}
\een
The kernel parameters are
\bea
\label{eq:tc15}
a&=&\frac{(z-1)}{4}\,u+\frac{z+1}{8\,z\,(z-1)\,u}-\frac{z^2+5\,z-2}{32\,z^2\,(z-1)^3\,u^3},\nonumber\\
b&=&\frac{(3\,z-1)^2}{16\,z^3}\,u,\\
&&-\frac{(3\,z-1)\,(18\,z^4-46\,z^3+31\,z^2-8\,z+1)}{64\,z^6\,
(z-1)^2\,u},\nonumber\\
\nu_f&=&2\,u\,z+\frac{4\,z^2-4\,z+1}{4\,z^2\,(z-1)\,u}\nonumber\\
&&-\frac{(2\,z-1)\,(8\,z^4-20\,z^3+26\,z^2-7\,z+1)}{64\,z^5\,(z-1)^3\,u^3}.\nonumber
\eea
Eqs. (\ref{eq:tc13}) and (\ref{eq:tc15}) are used to plot Figs. \ref{fig:comparison_largeu_z} and \ref{fig:comparison_weak_CT} of the main text.

\ssec{Dissociative limit}
\label{app:wavefunctions}
We analyse here the states and charge response in the dissociative limit, e.g.: $t\rightarrow 0$. Within the notation 
followed in this article, this means $z,\,u\rightarrow\infty$. We find that the many-body and KS ground states match in the dissociative CT regime, but are very different in the dissociative MH regime.
We start with the Kohn-Sham response.

\sssec{Kohn-Sham response}
The KS potential in the MH regime is zero, $x_s=0$. Therefore, $r_s=1$ and the wave-function coefficients $c_{s,\pm}=1/\sqrt{2}$.
Hence the KS HOMO / LUMO wavefunctions are bonding / antibonding orbitals
\bea 
|\phi^\pm\rangle=\frac{|1\rangle\mp |2\rangle}{\sqrt{2}},~~~(\rm{MH})
\eea
with energies $\mp t$.
As a consequence, the three singlet two-particle states are
\bea
|\Phi_0\rangle&=&\frac{|12\rangle+|21\rangle+|11\rangle+|22\rangle}{2},\nonumber\\
|\Phi_1\rangle&=&\frac{|11\rangle-|22\rangle}{\sqrt{2}},~~~(\rm{MH})\\
|\Phi_2\rangle&=&\frac{|12\rangle+|21\rangle-(|11\rangle+|22\rangle)}{2}.\nonumber
\eea
We also find that the excitation frequencies and weights are $\nu_s=1$ and $\nu_d=2$,
$W_s=2$, $W_d=0$. Finally, the KS charge response coefficients are $a_s=c_s=1/4$.

In contrast, the KS  potential in the CT regime is $x_s=x-u>0$.  Therefore $r_s=x-u$, the KS HOMO / LUMO
are 
\ben
|\phi^-\rangle=|1\rangle,\,\,\,|\phi^+\rangle=|2\rangle,~~~(\rm{CT})
\een
with energies $\mp(x-u)/2$. 
the singlet two-particle KS eigenstates in the dissociative CT regime are
\bea
|\Phi_0\rangle&=&|11\rangle,\nonumber\\
|\Phi_1\rangle&=&\frac{1}{\sqrt{2}}\,\left(|12\rangle+|21\rangle\right),~~~(\rm{CT})\\
|\Phi_2\rangle&=& |22\rangle.\nonumber
\eea
The excitation frequencies are $\nu_s=x-u$, $\nu_d=2\,\nu_s$, and the weights are
$W_s=2/(x-u)$ and $W_d=0$. The coefficients of the response function are
$a_s=(x-u)/4$ and $c_s=(x-u)^3/4$.
The results of this section are also used in the discussion in Subsection \ref{sec:tp33}.

\sssec{Many-Body response}
The eigenstates in the symmetric limit are
\ben
\label{eq:tc7}
|\Psi_0\rangle = \frac{|12\rangle+|21\rangle}{\sqrt{2}},\,\,|\Psi_{1,2}\rangle=\frac{|11\rangle \mp|22\rangle}{\sqrt{2}},
\een
while in the MH regime are
\ben
|\Psi_0\rangle=\frac{|12\rangle+|21\rangle}{\sqrt{2}},\,\,|\Psi_1\rangle = |11\rangle,\,\,|\Psi_2\rangle = |22\rangle.
\een
The overlap between the exact and KS ground state wavefunctions in the MH regime is
\ben
\langle\Psi_0|\Phi_0\rangle=\frac{1}{\sqrt{2}}.~~~(\rm{MH})
\een
The states $|\Psi_0\rangle$ and $|\Psi_1\rangle$ swap their nature at around $z=1$,
so in the CT regime the states are
\ben
|\Psi_0\rangle =  |11\rangle,\,\,|\Psi_1\rangle = \frac{|12\rangle+|21\rangle}{\sqrt{2}},\,\,|\Psi_2\rangle = |22\rangle.
\een
The overlap between the exact and KS ground state wavefunctions in the MH regime is 1.
We analyze only the MH regime from now on because the CT formulas are rather cumbersome and are not used in our interpolation. We find that the excitation frequencies and weights are
\ben
 \nu_{1,2}=u\mp x,\,\,\,\, W_i=\frac{2}{\nu_i^2}.~~~{\rm (MH)}
\een
Then, the kernel parameters are
\bea
a&=&\frac{1}{2\,\nu_3}=\frac{u^2-x^2}{8\,u},\,\,b=\frac{(u^2-x^2)^2}{2\,u\,(u^2+3\,x^2)},~~{\rm (MH)}\\
c&=&\frac{\nu_1\,\nu_2}{2\,\nu_4}=\frac{(u^2-x^2)^3}{8\,u\,(u^2+3x^2)},\,\,\,\nu_f^2=\frac{\nu_1\,\nu_2\,\nu_3}{\nu_4}=u^2+3\,x^2.\nonumber
\eea
Eq. (\ref{eq:tc7}) is used in Eq. (\ref{eq:tp4}) of the main text. The results of this section are also used in the discussion in Section \ref{sec:tp4} and Subsection \ref{sec:tp33}.

\end{document}